\documentclass[aps,prd,preprint,superscriptaddress,showpacs]{revtex4}
\usepackage{graphicx}

\usepackage{ulem}
\usepackage{color}
\definecolor{My_red}        {cmyk}{0.00,1.00,1.00,0.20}

\newcommand{\bmat}{\left(\begin{array}}
\newcommand{\emat}{\end{array}\right)}
\newcommand{\beq}{\begin{equation}}
\newcommand{\eeq}{\end{equation}}
\newcommand{\wt}{\widetilde}

\def\bwt{\begin{widetext}}
\def\ewt{\end{widetext}}
\def\be{\begin{equation}}
\def\ee{\end{equation}}
\def\bea{\begin{eqnarray}}
\def\eea{\end{eqnarray}}
\def\bean{\begin{eqnarray*}}
\def\eean{\end{eqnarray*}}
\def\bary{\begin{array}}
\def\eary{\end{array}}
\def\bit{\begin{itemize}}
\def\eit{\end{itemize}}

\def\su5u1{SU(5) \times U(1)}
\def\fsu5u1{SU(5) \times U(1)'}
\def\so10{SO(10)}
\def\sq20{SO(10) \times SO(10)}

\def\bwt{\begin{widetext}}
\def\ewt{\end{widetext}}
\def\be{\begin{equation}}
\def\ee{\end{equation}}
\def\bea{\begin{eqnarray}}
\def\eea{\end{eqnarray}}
\def\bean{\begin{eqnarray*}}
\def\eean{\end{eqnarray*}}
\def\bary{\begin{array}}
\def\eary{\end{array}}
\def\bit{\begin{itemize}}
\def\eit{\end{itemize}}

\def\su5u1{SU(5) \times U(1)}
\def\fsu5u1{SU(5) \times U(1)'}
\def\so10{SO(10)}
\def\sq20{SO(10) \times SO(10)}

\usepackage[centertags]{amsmath}
\usepackage{amssymb}

\begin{document}

\title{The Supersymmetric Standard Models with Decaying and Stable Dark Matters}

\author{Xin Gao}

\affiliation{Key Laboratory of Frontiers in Theoretical Physics, 
      Institute of Theoretical Physics, Chinese Academy of Sciences, 
Beijing 100190, P. R. China }

\author{Zhaofeng Kang}

\affiliation{Key Laboratory of Frontiers in Theoretical Physics, 
      Institute of Theoretical Physics, Chinese Academy of Sciences, 
Beijing 100190, P. R. China }

\author{Tianjun Li}

\affiliation{Key Laboratory of Frontiers in Theoretical Physics, 
      Institute of Theoretical Physics, Chinese Academy of Sciences, 
Beijing 100190, P. R. China }

\affiliation{George P. and Cynthia W. Mitchell Institute for
Fundamental Physics, Texas A$\&$M University, College Station, TX
77843, USA }

\date{\today}

\begin{abstract}

We propose two supersymmetric Standard Models (SMs) with decaying
and stable dark matter (DM) particles. To explain the SM fermion 
masses and mixings and have a heavy decay DM particle $S$,
we consider the Froggatt-Nielsen mechanism by introducing an
anomalous $U(1)_X$ gauge symmetry. Around the string scale,
the $U(1)_X$ gauge symmetry is broken down to a $Z_2$ symmetry
under which $S$ is odd while all the SM particles are even.
$S$ obtains a vacuum expectation value around the TeV scale,
and then it can three-body decay dominantly to the 
second/third family of
the SM leptons in Model I and to the first family 
of the SM leptons in Model II.
Choosing a benchmark point in the constrained minimal 
supersymmetric SM with exact R parity,  we show that the 
lightest neutralino DM is consistent with the CDMS II experiment.
Considering $S$ three-body decay and choosing suitable parameters,
we show that the PAMELA and Fermi-LAT experiments 
and the PAMELA and ATIC experiments can be explained
in Model I and Model II, respectively.

\end{abstract}

\pacs{12.60.Jv, 14.70.Pw, 95.35.+d}

%\preprint{MIFP-10-nn}

\maketitle

\section{Introduction }

It is well known that supersymmetry privides an elegant solution to
 gauge hierarchy problem in the Standard Model (SM). In the
Minimal Supersymmetric Standard Model (MSSM), gauge coupling
unification can be realized, which give us the important hint 
of Grand Unified Theory (GUT). In addition, in the supersymmetric
SMs, we can define a $Z_2$ symmetry called $R$ parity under which
the SM particles are even while their supersymmetric partners are
odd. With $R$ parity, we can avoid the dimension-four 
proton decay problem and evade the stringent constraints from the
electroweak precision data naturally. Interestingly, the lightest
supersymmetric particle (LSP) is stable due to the $R$ parity, and then
can be the dark matter (DM) candidate. For example, 
the lightest neutralino can be a viable cold DM candidate, which
can give us the correct relic density as well.

During the last two years, there were quite a few very interesting
DM experiments from indirect and direct dections. 
The ATIC~\cite{Chang:2008zz} and PPB-BETS~\cite{Torii:2008} 
collaborations have reported the measurements of 
cosmic ray (CR) electron/positron 
spectra at energies  up to $\sim 1$~TeV.
These data show an obvious excess over
the expected background in the energy ranges $\sim 300-800\,\textrm{GeV}$
and $\sim 500-800\,\textrm{GeV}$, respectively. In the mean time, the PAMELA
collaboration also released their first CR measurements of
the positron fraction~\cite{Adriani:2008zr} and the $\bar{p}/p$
ratio \cite{Adriani:2008zq}. Although  the $\bar{p}/p$
ratio is consistent with the astrophysical  expectation from
the interactions between the CR nuclei and interstellar
medium, the  positron fraction indeed
shows a significant excess for energies above $10\,\textrm{GeV}$ up to $\sim
100\,\textrm{GeV}$, compared to the background predicted by the 
conventional CR propagation models. Later, the Fermi-LAT collaboration 
has released data on the measurement of the electron/positron spectrum from 
20 GeV to 1 TeV with unprecedented precision~\cite{Abdo:2009zk}, 
and the HESS collaboration has released the data on  the measurements 
of electron/positron spectrum  from 
340 GeV to 700 GeV~\cite{Aharonian:2009ah}, complementing their 
earlier measurements from 700 GeV to 5 TeV~\cite{HESS:2008aa}. 
For simplicity, we will denote the Fermi-LAT collaboration as FERMI
collaboration in the following.
Although the corresponding data from the ATIC and the FERMI/HESS
experiments are not fully consistent, it was shown that the
DM models, where the DM particles annihilate or decay
 dominantly to the SM leptons, can explain these experiments 
by choosing the suitable DM particle
mass  and the proper final state particles.

 To explain the PAMELA/ATIC experiments or the PAMELA/FERMI/HESS
experiments from DM annihilations, we know that a large boost factor
about 100-1000 is needed. However, from
astrophysics, the N-body simulation shows that
 the boost factor from DM substructure can
never be larger than 10~\cite{Lavalle:1900wn}. To solve this problem, 
one can consider the Sommerfield 
enhancement~\cite {Hisano:2003ec, ArkaniHamed:2008qn, Nomura:2008ru}
or Breit-Wigner resonant 
enhancement \cite{Ibe:2008ye}. Alternatively,
we can also consider the non-thermal
dark matter production so that the DM annihilation
cross section can be large~\cite{Bi:2009am}. In addition, 
 if the DM particle is not absolutely stable and can decay dominantly to
leptons, we can explain these experiments for the DM lifetime at the 
order $\tau\sim 10^{25}-10^{27}s$~\cite{Chen:2008dh, Yin:2008bs,
Arvanitaki:2008hq, Ibarra:2009dr, Nardi:2008ix}.
In particular, 
 in the supersymmetric Standard Models, 
the LSP neutralino  cannot explain 
the PAMELA/ATIC experiments or the PAMELA/FERMI/HESS
experiments unless it can decay due to the suitable $R$-parity violation
dimension-four operators. 
Furthermore, to fit the PAMELA and ATIC data via the 
Markov Chain Monte Carlo (MCMC) technique, one found
that the DM mass is about 700 GeV for annihilation and 1.4 TeV
for decay, and the favored final state is $e^+e^-$~\cite{Liu:2009sq}. 
And to fit the PAMELA, FERMI, and HESS data, one found that 
 the DM mass is about 2 TeV for annihilation and 4 TeV
for decay, and the favored final states are the combination 
of  $\mu^+ \mu^-$ and $\tau^+ \tau^-$ since the electron/positron spectra
in the  FERMI and HESS experiments are softer than these in
the ATIC and PPB-BETS experiments~\cite{Liu:2009sq}. 
Also, the HESS obervation
of the Galactic center gamma rays gives  strong constraint
on the annihilation DM scenario while gives much weaker constraint
on the decay DM scenario. Thus, it favors the decay DM~\cite{Liu:2009sq}.

Recently, the Cryogenic Dark Matter Search (CDMS) collaboration
has observed two candidate DM events 
in the CDMS II experiment~\cite{Ahmed:2009zw}. 
The recoil energies for these two events are  12.3 keV 
and 15.5 keV, respectively, and the data set 
an upper limit on  the DM-nucleon elastic-scattering spin
independent cross section around $10^{-8}-10^{-7}$ pb.
Because the probability of observing two or more background events is
$23\%$, the CDMS II results cannot be a statistically
significant evidence for DM interactions, but these two 
events can not be rejected as signal. In particular, the favored
DM mass from the CDMS II data is about 100 GeV. 
Later, the CDMS II results have been studied extensively
in various DM models~\cite{Kadastik:2009ca}. 
Interestingly, the CDMS II experiment can be explained in 
the supersymmetric Standard Models
where the LSP neutralino is DM.

In short,  if  the PAMELA, ATIC, FERMI, and HESS experiments indeed
observed the DM annihilations or decays, the corresponding
DM particle is heavy around a few TeVs. And if the two 
events observed by the CDMS II experiment are DM signals, the 
corresponding DM particle is light around 100 GeV. Therefore,
there may exist at least two DM particles in the  Nature.
In fact, in almost all the previous DM models, the nearly 
universal implicit assumption is that there is  
one and only one DM particle.
However, we cannot prove this implicit assumption, and then we cannot
ingore the possibility of multicomponent DM~\cite{MC-DMs, Hur:2007ur}. 

In this paper, we propose two supersymmetric Standard Models
with decaying and stable DM particles. To avoid the proton decay
problem and evade the stringent constraints from the
electroweak precision data, we assume that $R$ parity is
not violated. In our models, we require that
the LSP neutralino be the stable DM particle with mass around 100 GeV
and the LSP neutralino-nucleon scattering cross section
be about $10^{-8}$ pb. Thus,
 we can explain the CDMS II experiment. We also assume that
the supersymmetry breaking scale is still below  1 TeV,
and then we can solve the gauge hierarchy problem without
fine-tuning.
To explain the PAMELA, ATIC, and FERMI experiments, 
we introduce a DM particle $S$ with mass around
a few TeVs. To produce the SM fermion masess and mixings and 
have the heavy decay DM particle $S$, we consider the 
Froggatt-Nielsen (FN) mechanism~\cite{Froggatt:1978nt}. 
We introduce an anomalous $U(1)_X$ gauge symmetry 
whose anomaly is cancelled by the Green-Schwarz 
mechanism~\cite{Green:1984sg}. Especially,
the $U(1)_X$ gauge symmetry is broken 
down to the $Z_2$ symmetry around
the string scale under which $S$ is odd while all the SM
particles are even. Thus, $S$ can be a DM particle.
Similar to the discussions 
in Refs.~\cite{Ibanez:1994ig, Dreiner:2003yr, Harnik:2004yp,
Gogoladze:2007ck},
the SM fermion masses and mixings can be generated as well.
With a pair of heavy vector-like particles that are SM singlets
and have $U(1)_Y$ charges $\pm 1$ respectively, we obtain
that the leading Yukawa coupling
terms between $S$ and the SM particles are $S^2 H_d L_i E^c_j$ 
where $H_d$ is the Higgs field, and $L_i$ and $E^c_j$ are the
$i$-th family of the SM lepton doublet and $j$-th family
of the right-handed charged lepton, respectively. 
In Model I, we have
$(i,~j)=(2,~2)$ and $(3,~2)$, while in Model II, we have
$(i,~j)=(1, ~1)$.
After $S$ obtains a vacuum expectation value (VEV) at the
TeV scale, $S$ can decay dominantly to 
the second/third  family of the SM leptons via dimension-six operators 
(three-body decay) in Model I, and to the first family of the 
SM leptons in Model II. To realize our idea,
we present a benchmark point from the constrained minimal 
supersymmetric Stanard Model (CMSSM). The lightest 
neutralino ${\wt N}_1^0$ 
contributes to part of the whole DM relic density, {\it i.e.}, 
$\Omega_{{\wt N}_1^0 }h^2\approx 0.08$. The LSP
neutralino-nucleon elastic-scattering spin
independent cross section is about $5\times 10^{-9}$ pb.
Thus, this benchmark point is 
consistent with the CDMS II results. In addition,
for the $S$ lifetime  about  
$\tau\sim 10^{25}-10^{27}s$, we can explain
the PAMELA, FERMI and CDMS II experiments in Model I with
$S$ mass 3 TeV, and explain  the PAMELA, ATIC and 
CDMS II experiments in Model II
with $S$ mass 1.8 TeV.

The paper is organized as follows. In Section II, we explain
the SM fermion masses and mixings as well as the leading
Yukawa terms $S^2L_iH_dE^c_j$ via the FN mechanism.
 In Section III, we  present a benchmark point in the CMSSM 
parameter space, and we
discuss the DM particle $S$ three-body cascade decay.
We fit the PAMELA and FERMI data and the PAMELA and ATIC
data in Section IV.  Section V is our discussion and conclusions.
We present more technical details in Appendices A, B, and C.

\section{Two Supersymmetric Standard Models}

In this Section, we will present two supersymmetric Standard Models with
decaying and stable DM particles. To have a decay DM particle,
we introduce a SM singlet field $S$ and a new $Z_2$ symmetry.
Under this $Z_2$ symmetry, $S$ is odd while all the 
SM particles are even. 
Thus, $S$ can be a DM particle. Because the global discrete
symmetry can be broken by the quantum gravity effects, it is
natural to have the remnant discrete symmetry from an extra $U(1)'$ gauge
symmetry~\cite{Ibanez:1991hv, Dreiner:2005rd}.  In  particular, 
in the $U(1)'$-extended Minimal 
Supersymmetric Standard Model (UMSSM)~\cite{Cvetic:1997ky} 
where we can solve the 
$\mu$ problem and avoid proton decay, we indeed have a new DM
candidate if the $U(1)'$ is broken down to a $Z_2$ 
symmetry~\cite{Hur:2007ur, Lee:2008pc}.

In our models, this $Z_2$ symmetry
is the residual symmetry after the anomalous $U(1)_X$
gauge symmetry breaking, ${\it i.e.}$, $Z_2$ is a subgroup of $U(1)_X$.
After supersymmetry is broken
at the TeV scale,  the supersymmetry breaking soft masses may make 
 $m_{S}^2$ negative and then the scalar component
of $S$ obtains a VEV around the TeV scale. 
We assume that the lightest state in the supermultiplet $S$
is its scalar component, which is still denoted by $S$ in this 
paper. In short, $S$ is a DM particle which can decay 
after $Z_2$ symmetry breaking. 
In the concrete model
building, we may need a singlet sector with additional particles
and interactions for two kinds of reasons: (1) We have to
 stabilize the VEV of $S$ at the TeV scale; (2) We need to generate
the correct $S$ relic density from thermal productions by considering
the additional $U(1)'$ gauge symmetry at the TeV scale~\cite{Hur:2007ur}, 
or from non-thermal productions by considering the cosmic string at
intermediate scale~\cite{Bi:2009am} or the 
extra vector-like particles at the intermediate
scale~\cite{Arvanitaki:2008hq}. Similar to the 
Ref.~\cite{Arvanitaki:2008hq},
we concentrate on the DM phenomenology in this paper,
and then the detailed discussions of the singlet sector
and $S$ relic density
 are out of the scope of our current paper.

To explain the PAMELA, ATIC, FERMI and HESS experiments,
$S$ must decay dominantly to the SM leptons. To select
the suitable final state particles, we consider the 
Froggatt-Nielsen mechanism~\cite{Froggatt:1978nt}, which is
an elegant way  to explain the SM fermion masses and
mixings. We consider an anomalus $U(1)_X$
gauge symmetry, whose anomaly can 
be cancelled via  the Green-Schwarz mechanism~\cite{Green:1984sg}.
In the string model building, we indeed have
at least one anomalous $U(1)_X$ gauge symmetry~\cite{Gogoladze:2007ck}.

To break the $U(1)_X$ gauge symmetry, we introduce a flavon field
$A$ with $U(1)_X$ charge $-1$. Because supersymmetry must be preserved
close to the string scale, $A$ can acquire a VEV
so that the $U(1)_X$ D-flatness can be realized.  
It was shown~\cite{Dreiner:2003yr} that
\begin{equation}
0.171 \le \epsilon\equiv{\frac{\langle A \rangle}{M_{\rm Pl}}} \le 0.221
~,~\,
\end{equation}
where $M_{\rm Pl}$ is the reduced Planck scale.
Interestingly, $\epsilon$ is about the size of the Cabibbo angle.
Let us explain our convention. We denote the
 SM left-handed quark doublets,
right-handed up-type quarks, right-handed down-type
quarks, left-handed lepton doublets, right-handed
neutrinos, and right-handed charged leptons as 
$Q_i$, $U^c_i$, $D^c_i$, $L_i$, $N^c_i$, and $E^c_i$, respectively.
Also, there is one pair of Higgs doublets $H_u$ and $H_d$
in the supersymmetric Standard Models. 
Moreover, we introduce a pair of vector-like particles $E'$ and
$\overline{E}'$ whose quantum numbers under 
$SU(3)_C\times SU(2)_L\times U(1)_Y$ are 
$(\mathbf{1}, \mathbf{1}, \mathbf{-1})$
and $(\mathbf{1}, \mathbf{1}, \mathbf{1})$, respectively.
In addition, the $U(1)_X$ charges for all the particles
in our models are denoted by appropriate subscripts, for example,
for a generic particle $\phi$, its $U(1)_X$ charge is $X_{\phi}$.
In our models, we choose the $U(1)_X$ charges
for the SM fermions and Higgs fields as follows
\begin{align}
X_{Q_1}=3,\quad X_{Q_2}=2,\quad X_{Q_3}=0,\quad
X_{U^c_1}=5, \quad X_{U^c_2}=2,\quad  X_{U^c_3} =0, \cr
X_{D^c_1}=1, \quad X_{D^c_2}=0,  \quad X_{D^c_3}=0, \quad 
X_{L_1}=1,\quad  X_{L_2}=0,\quad X_{L_3}=0, \cr
X_{E^c_1}=3, \quad X_{E^c_2}=2,\quad X_{E_3^c}=0, \quad 
X_{H_u}=0, \quad X_{H_d}=0. ~~~~~~~~~~~
\end{align}

The superpotential for the SM fermion Yukawa couplings is 
\begin{align}
 W \supset &\left(A\over M_{\rm Pl}\right)^{X_{H_u}+X_{Q_i}+X_{U^c_j}}Q_iH_uU^c_j+
\left(A\over M_{\rm Pl}\right)^{X_{H_d}+X_{Q_i}+X_{D^c_j}}Q_iH_dD^c_j\cr
+&\left(A\over
M_{\rm Pl}\right)^{X_{H_d}+X_{L_i}+X_{E^c_j}}L_iH_dE^c_j+\left(A\over
M_{\rm Pl}\right)^{X_{H_u}+X_{L_i}+X_{N^c_j}}L_iH_uN^c_j~,~
\end{align}
where $i,~j=1,~2,~3$, and all the coefficients  are 
assume to be order one in the above superpotential.
Thus, we obtain the  SM fermion Yukawa coupling matrixes 
$Y_u$, $Y_d$, $Y_e$ and $Y_{\nu}$ respectively for
up-type quarks, down-type quarks, charged leptons and
active neutrinos
\begin{align}
Y_u\sim \left(
\begin{array}{ccc}    \epsilon^8  &\epsilon^5 &
\epsilon^3\\
  \epsilon^7  &   \epsilon^4   & \epsilon^2   \\
  \epsilon^5 &\epsilon^2  &  1
\end{array} \right),\quad Y_d\sim Y_e^T \sim \left(
\begin{array}{ccc}    \epsilon^5  &\epsilon^4 &
\epsilon^4\\
  \epsilon^4  &   \epsilon^3   & \epsilon^3   \\
  \epsilon^2 &\epsilon^1  &  \epsilon^1
\end{array} \right),\quad Y_{\nu}\sim \left(
\begin{array}{ccc}    \epsilon^2  &\epsilon &
\epsilon\\
  \epsilon  &  1  & 1   \\
  \epsilon & 1  &  1
\end{array} \right)~,~
\end{align}
where $T$ is transpose.
We can show that the observed SM fermion masses 
and mixings can be generated~\cite{Gogoladze:2007ck, Dreiner:2003yr},
and $\tan\beta=\langle H_u^0\rangle/\langle H_d\rangle^0$  is $25$. 

In the Model I, we would like to explain the PAMELA and FERMI
 experiments. We choose the following $U(1)_X$
charges for $S$,
$E'$ and $\overline{E'}$
\begin{align}
X_S=3/2, \quad X_{E'}=-5, \quad X_{\overline{E}'}=5. 
\end{align}
Note that $A$ has $U(1)_X$ charge $-1$, the $U(1)_X$ charges for the 
SM particles are integers, while the $U(1)_X$ charge for $S$ is
half integer, thus, the $U(1)_X$ gauge symmetry is broken down
to a $Z_2$ symmetry after $A$ obtains a VEV. In particular,
under this $Z_2$ symmetry, only $S$ is odd while all the other 
SM particles are even.
Then the leading Yukawa coupling terms between $S$ and 
the SM particles in the superpotential are
\begin{align}
 W \supset & \left(A\over M_{\rm Pl}\right)^5
 H_d L_k \overline{E'}  + \left(S^2\over
M_{\rm Pl}\right) E^c_2 E' + M_V \overline{E'} E' ~,~
\label{SP-Model-I}
\end{align}
where $k=2,~3$, and $M_V$ is the vector-like particle
mass around the $10^{13}$ GeV. In particular, the 
superpotential mass terms $ E_i^c E'$ and $A^n E_i^c E'$ 
are forbidden due to the
holomorpic property of superpotential.
Because $E'$ and $\overline{E'}$ are heavy, we should
integrate them out below their mass scale. From the above
superpotential, we obtain 
from equations of motion (EOMs)
for $E'$ and $\overline{E'}$
\begin{align}
E'~=~ -{{H_d L_k}\over {M_V}}~,~~~  
\overline{E'} ~=~ -{{S^2 E_2^c}\over {M_V M_{\rm Pl}}}~.~
\label{EOM-Model-I}
\end{align}
Using the above EOMs for $E'$ and $\overline{E'}$,
 we obtain that the superpotential in Eq.~(\ref{SP-Model-I})
becomes
\begin{align}
 W \supset~ - \epsilon^5 \left(S^2\over
{M_{\rm Pl}M_V}\right)  H_d L_k E^c_2~.~\,
\end{align}
To simplify the discussions, we assume that the coefficient of
$S^2 H_d L_2 E^c_2$ is about three times larger than that of 
$S^2 H_d L_3 E^c_2 $. Thus, we will concentrate on the term
$S^2 H_d L_2 E^c_2$ in the following discussions.

In the Model II, we want to explain the PAMELA
and ATIC experiments. We choose the $U(1)_X$
charges for $S$,
$E'$ and $\overline{E'}$ as follows
\begin{align}
X_S=1/2, \quad X_{E'}=-4, \quad X_{\overline{E}'}=4. 
\end{align}
Similar to the Model I, the $U(1)_X$ gauge symmetry is broken down
to a $Z_2$ symmetry after $A$ obtains a VEV. Under this $Z_2$ symmetry,
only $S$ is odd while all the other SM particles are even.
Then the leading Yukawa coupling terms between $S$ and 
the SM particles in the superpotential are
\begin{align}
 W \supset & \left(A\over M_{\rm Pl}\right)^4
H_d L_1 \overline{E'}  + \left(S^2\over
M_{\rm Pl}\right) E^c_1 E' + M_V \overline{E'} E' ~.~
\end{align}
Similar to the Model I,  integrating out the vector-like particles
$\overline{E'}$ and $E'$, we obtain 
\begin{align}
 W \supset~ - \epsilon^4 \left(S^2\over
{M_{\rm Pl}M_V}\right)  H_d L_1 E^c_1~.~\,
\end{align}

In this paper, we will define $M_*^2=  M_{\rm Pl}M_V/\epsilon^{n}$
where $n$ is equal to $5$ in Model I and $4$ in Model II. 
With $M_V$ around $10^{13}$ GeV, we obtain that $M_*$ is
around $10^{17}$~GeV. To get $M_V$ around $10^{13}$ GeV,
we can introduce additional global $U(1)''$ symmetry and
a SM singlet field $S'$ which breaks the  $U(1)''$ symmetry. 
The $U(1)''$ charges for $L_i$, $E_i^c$, $\overline{E'}$, $S$,
and $S'$ are $1$, $-1$, $-1$, $1/2$ and $1$, while the
$U(1)''$ charges for all the other particles are zero.
Therefore, all the previous terms in the superpotential 
including the SM fermion Yukawa coupling terms  are
invariant except the term $M_V \overline{E'} E'$. The vector-like
mass term for $\overline{E'}$ and $E'$ can be generated by
the following superpotential term after the $U(1)''$ symmetry breaking
\begin{align}
W ~\supset~ S' \overline{E'} E'~.~\,
\end{align}
Assuming that $S'$ acquires a VEV around $10^{13}$ GeV,
we obtain that $M_V$ is around $10^{13}$ GeV.

By the way, the other Yukawa coupling terms 
among $S$, the SM fermions and Higgs fields might have the prefactor
$\left(S^2\over M_{\rm Pl}^2\right)^n
\left(A\over M_{\rm Pl}\right)^m$(...) ($n\geq 1$ or
$m\geq1$). Thus, these Yukawa coupling terms are 
supressed by at least the Planck scale square and then are negligible
since $S$'s VEV is around the TeV scale. In this paper, we only consider
the minimal K\"ahler potential. In general, the dimension-six
operators of the form $SS^\dagger \phi \phi^\dagger/M^2_{\rm Pl} $ 
 cannot be forbidden  by any symmetry where $\phi$ denotes the SM
fermions and Higgs fields.  These dimension-six operators
can  induce DM two-body decay through derivative 
couplings~\cite{Arvanitaki:2008hq}. However, their contributions to
the CR are also ignorable since these operators are 
suppressed by the Planck scale square as well.

\section{Decay and Stable Dark Matters \label{1}}

\subsection{The CMSSM Benchmark  Point for CDMS II Experiment}\label{massspectrum}

The CMSSM with the LSP neutralino as DM
  has been studied extensively before.
Typically, the LSP neutralino with mass about tens of GeV is 
in tension with the lightest CP-even Higgs boson mass whose low bound from the
LEP is 114 GeV. Especially, we require that the LSP neutralino relic
density be small about half of the total DM relic density, {\it i.e.}
 $\Omega_{{\wt N}_1^0} h^2\sim 0.06$. In this paper,
 we do not scan all the viable parameter space.  We only consider
a CMSSM benchmark point that satisfies all the constraints. 
We choose the following five free parameters at the GUT scale 
\begin{align}
m_0=310{\rm GeV},\quad m_{1/2}=250{\rm GeV},\quad A_0=-1040,\quad
\tan\beta=30,\quad {\rm sign}(\mu)=+1.
\end{align}
As expected, this benchmark point is in the
coannihilation region~\cite{Edsjo:1997bg} with small mass difference
between the light stau ${\wt \tau_1}$ and LSP neutralino, {\it i.e.}, 
$m_{\wt \tau_1}-m_{{\wt N}_1^0}\leq 8$ GeV. 
Such benchmark point is not interesting previously since the LSP 
neutralino relic 
density is smaller than the observed whole DM relic density. However, 
it is fine in our models since we have two DM particles.
With MicrOMEGAs 2.0~\cite{Belanger:2006is}, we obtain the 
LSP neutralino relic density
\begin{align}
\Omega_{{\wt N}_1^0}h^2\approx 0.08~.~
\end{align}
Moreover, the mass of the LSP neutralino  is about $ 101.6$ GeV. And
the spin-independent  cross section between the LSP neutralino 
and nucleon is about
$\sigma_{SI}\approx 5\times 10^{-9}$ pb. Although this cross
section is a little bit small, we can still explain
the CDMS II experiment due to the uncertainties of
the  QCD effects in the  calculations.

It is necessary to address the mass spectrum of this benchmark 
point in details for the following calculations. The lightest neutralino
$\wt N_{1}$ is bino-like, while the heavy neutralinos $\wt N_{3,4}$
are Higgsino-like. Since  $\tan\beta$ is large, in the charged Higgs
boson system, $H_u^+$ is the major component of the Goldstone boson,
with the ratio to $(H_d^-)^*$ to be $\tan^2\beta:\,1$. So we can
treat $H_d^-$ as the charged Higgs boson $C^-$ approximately, 
{\it i.e.}, $H_d^-\approx C^-$. The CP-odd neutral Higgs
and charged Higgs mass matries are diagonlized by the same matrix.
Thus, the imaginary part of $H_u^0$ gives  the main component of
the Goldstone boson, while
 $A^0$ is mainly the imaginary part of $H_d^0$, {\it i.e.},
$A^0\approx{\rm {Im}}(H_d^0)$. Same conclusion applies to the
heavy CP-even Higgs boson $H^0$, {\it i.e.}, 
$H^0\approx {\rm{Re}} (H_d^0)$ because  $H^0$ is equal to 
$\cos\alpha {{\rm Re}(H_d^0)}+\sin\alpha {{\rm Re}(H_u^0)}$ with
$\alpha\approx-0.04$. As for the chargino system, the large $\mu$ term
implies that the charged Higgsino is the major component of 
heavy chargino $\wt C_2^-$.  Because we need to use the neutralino,
chargino and Higgs mixing matrices in the following discussions,
we calculate the relevant masses and mixings at low energy 
by SuSpect~\cite{Djouadi:2002ze}. For the
neutralino system, the four particles $(-i\wt B,-i\wt W_3,\wt
H_d^0,\wt H_u^0)$ are transformed into the mass eigenstates $(\wt
N_1^0,\wt N_2^0,\wt N_3^0,\wt N_4^0)$ by the unitary matrix $Z$
\begin{align}
\left( \begin{array}{cccc}-i\wt B\\-i\wt W_3\\\wt H_d^0\\\wt
H_u^0\end{array} \right)=\left(
\begin{array}{cccc}  1.00&     -0.02  &0.08  &
-0.02\\
  0.03  &    0.99   &  -0.16   &   0.06\\
  0.04 &-0.07  &  -0.70  &  -0.71\\
  0.07 &  -0.15 &   -0.69  &    0.70
\end{array} \right)\left( \begin{array}{cccc}\wt N_1^0\\\wt N_2^0\\\wt
N_3^0\\\wt N_4^0 \end{array}\right).
\end{align}
And the neutralino mass eigenvalues  are $m_{\wt
N_{1,2,3,4}}=101.6,\,196.2,\,571.0, \,577.0$~GeV, respectively.
The particles ($W_3^-$, $H_d^-$) and 
(${\rm Re}(H_d^0)$, ${\rm Re}(H_u^0)$) can be written respectively
in terms of their mass eigenstates as follows
\begin{align}
 \left(
\begin{array}{cc}\wt W_3^- \\ \wt H_d^-
\end{array}\right)=\left(
\begin{array}{cc}  -0.98&     0.22  \\
  0.22  &    0.98  \\
\end{array} \right)\left( \begin{array}{cc}\wt C_1^-\\ \wt C_2^-
\end{array}\right),\quad
\left( \begin{array}{cc}{\rm Re}(H_d^0)\\ {\rm Re}(H_u^0)
\end{array}\right)&=\left(
\begin{array}{cc}1.00  &     0.04  \\
  -0.04  &    1.00  \\
\end{array} \right)
\left( \begin{array}{cc}H^0\\h^0\end{array} \right),
\end{align}
with $m_{\wt C^-_{1,2}}=196.1,\, 578.5 $ GeV, and $m_{h^0,H^0}=
  117.90,\,510.81$ GeV. The Im$(H_d^0)$ and Im$(H_u^0)$ are not
given explicitly here, and 
the mass of the CP-odd Higgs field  $A^0$ is  $512.49$ GeV.

\subsection{DM $S$ Three-Body Decays}

In this paper, we will concentrate on the calculations in Model I.
For simplicity, we will not explain the calculations in Model II
since the calculations are not only similar but also simpler.
To analyze the DM $S$ primary decays, we write the operator in
the  components  explicitly
\begin{align}\label{decays}
{\cal{C}}_\mu S H_dL E^c+&c.c.= {\cal{C}}_\mu S H_d^0
EE^c-{\cal{C}}_\mu S H_d^-\nu E^c + c.c.\cr \supset& {\cal{C}}_\mu S (H_d^0
\mu_L\mu_R^\dagger+\wt H_d^0 \mu_L \wt \mu^*_R +\wt H_d^0 \wt \mu_L
\mu_R^\dagger)-{\cal{C}}_\mu S(H_d^-\nu_\mu \mu_R^\dagger+\wt
H_d^-\nu_\mu \wt \mu_R^*+\wt H_d^-\wt\nu_\mu \mu_R^\dagger) \cr &+
{\cal{C}}_\mu \wt S (H_d^0 \mu_L \wt \mu^*_R+H_d^0 \wt \mu_L \mu^*_R
+\wt H_d^0 \wt \mu_L \wt \mu^*_R+\wt H_d^0 \wt \mu_L \wt
\mu_R^\dagger) \cr &
 -{\cal{C}}_\mu \wt S(H_d^-\nu_\mu \wt
\mu_R^*+H_d^-\wt \nu_\mu  \mu_R^*+ \wt H_d^-\wt \nu_\mu \wt
\mu_R^*+\wt H_d^-\wt\nu_\mu \wt \mu_R^\dagger),
\end{align}
where ${\cal{C}}_\mu$ is equal to $\langle S\rangle/M_{*}^2$ times
the order one coefficient. If these states are transferred to the mass
eigenstates, further mixing factor must be included. After $H_d^0$
obtains a VEV, we shall have two-body and three-body $S$ decays.
Comparing to the two-body $S$ decays, the three-body 
$S$ decays have an extra factor $(M_D/\langle H_d^0\rangle)^2/(96\pi^2)$
where $m_D$ is the mass of the heavy DM particle $S$.
In our models, we choose $M_D=3$ TeV and $\tan\beta =30$, and then
we get $(M_D/\langle H_d^0\rangle)^2/(96\pi^2) \sim 282$,
Therefore, we only consider three-body $S$ decays
in this paper.

The heavy decay DM particle is the scalar component of $S$,
whose three-body  on-shell decays are given by the second
line of Eq.~(\ref{decays})
\begin{align}\label{onshelldecay}
S\rightarrow \wt N_{3,4} \mu \wt \mu_R,\quad\wt C_{1,2}\nu_\mu \wt
\mu_L ,\quad \wt C_{1,2} \mu \wt{\nu_\mu}_L,\quad A^0(H^0)
\mu\mu,\quad C^-\nu_\mu  \mu_R~.~\,
\end{align}
With the Eq.~(\ref{decayrate}) in the Appendix B, we obtain that all
these nine primary decay channels almost have the democratic rates except
the mixing factor  for each channel. Anyway, the total
decay rate is given by
\begin{align}
\Gamma_{total} ={1\over 128\pi^3} {\langle S\rangle^2m_D^2\over
M_{*}^4} \left(\sum_{I}\mathcal {C}_{I}^2R_{I}\right)\times
m_D,
\end{align}
where $I$ denotes  the $I$-th decay channel in
Eq.~({\ref{onshelldecay}}).
In our paper, the DM particle $S$  is so heavy that the phase space
suppressing factor $R_{I}$ is almost process independent and is
 a constant around 1/6 (see Eq.~(\ref{decayrate})). Then,
the scale of the DM $S$ lifetime is estimated to be
\begin{align}\label{life}
\tau_S \sim 2.7 \times 10^{26}\times \left(3 {\rm TeV}\over
m_D\right)^3 \left(M_{*}\over 10^{17}{\rm GeV}\right)^4
\left(3\,{\rm TeV} \over \langle S\rangle\right)^2\left({1}\over
\sum_I\mathcal {C}_{I}^2R_{I}\right)s.
\end{align}
The random coefficient of the operator has been set to be 1. Notice
that except the last factor, Eq.~(\ref{life}) is the so-called
$\tau_{eff}$ which is the inverse of $\Gamma_{eff}$ defined in
Eq.~(\ref{decayrate}).

\section{Cosmic Ray Anomalies}

Although the very large astrophysical uncertainties 
do exist in the calculations
of cosmic ray, we do not  want to scan all the
viable parameter space. We only want to demonstrate that 
with appropriate parameters, we can explain
the  PAMELA and FERMI experiments in Model I 
and explain the PAMELA and ATIC experiments in Model II.

\subsection{Cosmic $e^\pm$ Excess}

In the Micky Way dark halo, the DM $S$ decays into the SM
particles, which propagate to the solar system. The propogation of
 charged particles is described by  diffusion equation. For 
instance, the diffusion equation for  positron is
\begin{align}\label{pos}
{\partial \psi\over \partial t}-\nabla\cdot({K(\vec{x},E)\nabla
\psi}) -{\partial \over \partial
E}\left({b(E)\psi}\right)=q(\vec{x},E),
\end{align}
where $\psi(\vec{x},E)$ denotes the positron number density per unit
energy. Diffusion coefficient $K=K_0E^\delta$ is space independent.
The third term describes the energy loss of positron through the
synchrotron radiation and inverse Compton scattering, with loss
ratio  $b(E)=E^2/\tau_E$ where $\tau_E=10^{16}s$. $q(\vec{x},E)$ is
the positron source term, describing the positron  number density
injected at $\vec{x}$, per unit time and energy. 
For simplicity, it can be
expressed as the product of an astrophysics factor and  a particle
physics factor
\begin{align}
q(\vec{x},E)=\left({1\over \tau_S}\sum_{I,F} {dN_{I,F}^e\over
dE}B_{I,F}\right)\times {\rho(\vec{x})\over m_{D}},
\label{Q-Source}
\end{align}
where we introduce another index $F$  to 
distinguish the different states from  the  $I$-th 
decay channel. Moreover,
 the Milky Way DM density profile $\rho(\vec{x})$ in the
spherical coordinates is generically written as
\begin{align}
\rho(r)=\rho_\odot \left({r_\odot\over r}\right)^\gamma
\left(1+(r_\odot/r_s)^\alpha\over
1+(r/r_s)^\alpha\right)^{(\beta-\gamma)/\alpha}~,~\,
\end{align}
where $r_\odot=8.5$ kpc  is  the distance from  the solar system to the
Milky Way center, and  the DM profile density in the solar system
is set to be $\rho_\odot=0.3\,{\rm GeV\,cm}^{-3}$.  According to the
$N-$body simulation, it has three popular choices parameterized by
$(\alpha,\beta,\gamma)$. We use the NFW 
profile with $(1,3,1)$~\cite{Navarro:1996gj}, and choose the DM
central core $r_s=20$ kpc. By the way, the injected source term is
linearly  proportional to the decay DM  relic density. Therefore, 
 the fraction of the DM $S$ relic density $r_{DM}$ 
to the whole DM relic density should be
taken into account. For simplicity,  we absorb it into
the redefinition of $S$ lifetime by $\tau_S \rightarrow r_{DM} \tau_S$ without
disturbing any other terms.

The normalized energy spectrum distribution function ${dN_{I,F}^e/
dE}=({2/ m_D}){dN_{I,F}^e/ dx}$ describes the positrons from the specific
DM decay intermediate state $(I,F)$, with $(I,F)=\wt{
\mu}_L,\, A^0...$, and so on. And $B_{I,F} = B_I$ is the corresponding
branch ratio. In this paper, summing over all the channels (see
Appendix, with a little bit changed notation), we obtain the total
fragmentation function
\begin{align}{\label{total}}
{dN_e^S\over dx}
 =&{ \Gamma_{eff}\over \Gamma_{total}} \sum_{I,F}
\left(\mathcal {C}_{I,F}^2R_{I,F}\right)\times {1
\over R_{I,F}}{d\wt{N}\over dx_{I,F}}\otimes {dN\over dx^{I,F}_e },
\end{align}
where  $dN/dx_{I,F}$ and $1/R_{I,F}dN/dx_e^{I,F}$ 
respectively give
the fragmentation functions in the rest frame of the DM $S$ and
particle $({I,F})$. The former is analytically calculable, while the
later is obtained through PYTHIA~\cite{Sjostrand:2006za} simulation
(except for muon decay, we shall use the analytical formula). The
total spectrum is their convolution, whose explicit form can be
found in Eq.~({\ref{convolution}}). By the way, when we put 
Eq.~(\ref{total}) into Eq.~(\ref{Q-Source}), the $\tau_S$
 factor in the front of source term in Eq.~(\ref{Q-Source}) 
is cancelled by factor $\Gamma_{total}$ in Eq.~(\ref{total}).
So,  $\tau_{eff}$ or $\Gamma_{eff}$ is still  the free parameter in the
calculations of positron propagation.

Eq.~(\ref{pos}) can be approximately solved analytically. The final
positron flux observed at the Earth $\Phi_{e^+}(r_\odot,E)$ is
factorized into the convolution of $\mathcal {H}_e(E_s,E)$ which encodes
the whole astrophysical information and the inject spectrum
$dN^e_S/dE_s$
\begin{align}
\Phi_{e^+}(r_\odot,E)={\beta_{e^+}\over 4\pi}\kappa {\tau_E\over
E^2} \int_E^{E_{max}} \mathcal {H}_e(E_s,E)dE_s {dN_S^e\over dE_s}~,
\end{align}
where $\beta_{e^+}$ is the velocity of the observed positron and
$\kappa={\rho_\odot/(\tau_{eff}\, m_D)}$. The concrete  method to
calculate the halo functions 
for electron $\mathcal {H}_e(E_s,E)$ and for anti-proton 
$\mathcal {H}_{\bar p}(E)$ can be found 
in Ref.~\cite{Lavalle:1900wn,Delahaye:2007fr}.

 To compare the data, we should consider
the cosmic background.  It is believed that the astrophysical
$e^\pm$ source is mainly due to supernova explosions (primary $e^-$)
and the interactions between the CR nuclei  and light atoms in
interstellar medium (secondary $e^\pm$). They are usually
parameterized in the following form \cite{Baltz:1998xv}
\begin{align}
\Phi_{e^-}^{bk,pri}=&{0.16E^{-1.1}\over
1+11E^{0.9}+3.2E^{2.15}},\quad\Phi_{e^-}^{bk,sec}={0.7E^{0.7}\over
1+110E^{1.5}+600E^{2.9}+580E^{4.2}},\cr
~~~~~~~~~~~~\Phi_{e^+}^{bk,sec}=&{4.5E^{0.7}\over 1+650E^{2.3}+1500E^{4.2}},
\end{align}
where the unit is GeV$^{-1}$cm$^{-2}s^{-1}sr^{-1}$.
$\Phi_{e^-}^{bk,pri}$, $\Phi_{e^-}^{bk,sec}$, and $\Phi_{e^+}^{bk,sec}$
are the background fluxes for the primary electron, secondary electron,
and secondary positron, respectively. Then
 the positron fraction observed by the PAMELA experiment
and the normalized total electron/positron
fluxes observed by the FERMI and ATIC experiments
are respectively given by
\begin{align}
{\Phi_{e^+}\over\Phi_{e^+}+\Phi_{e^-}}&={\Phi_{e^+}^{bk,sec}+\Phi_{e^+}^{DM}
\over\kappa\Phi_{e^-}^{bk,pri}+\Phi_{e^+}^{bk,sec}+\Phi_{e^-}^{bk,sec}
+\Phi_{e^+}^{DM}+\Phi_{e^-}^{DM}}~,~\, \cr
 E^3_e(\Phi_{e^+}+\Phi_{e^-})&=E_e^3(\kappa\Phi_{e^-}^{bk,pri}
+\Phi_{e^+}^{bk,sec}+\Phi_{e^-}^{bk,sec}+\Phi_{e^+}^{DM}+\Phi_{e^-}^{DM})~,~
\end{align}
where $\kappa\leq1$ is a  parameter which includes the
uncertainties of the primary background electron production.

\begin{figure}[htb]
\begin{center}
\includegraphics[width=6.5in]{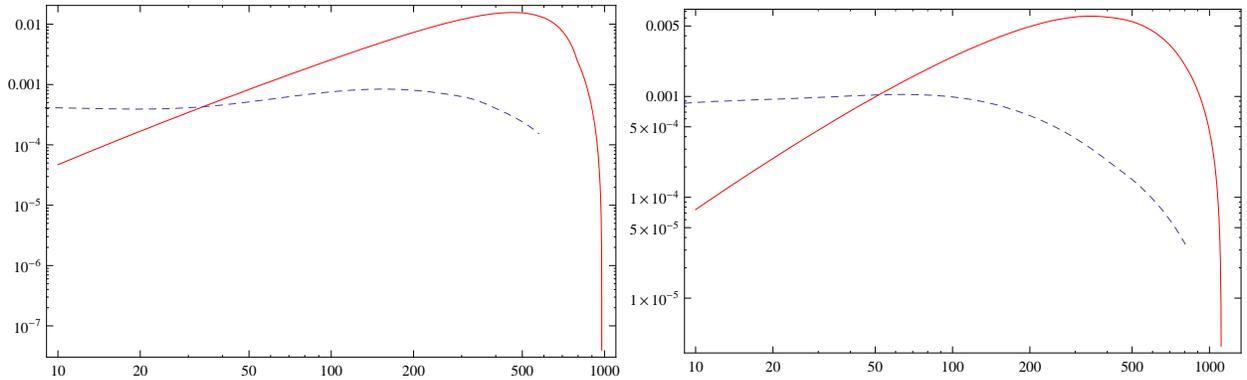}
\end{center}
\caption{\label{compare} The comparison between  
the contributions to total $e^\pm$ fluxes from 
$\mu$ in Model I (or $e$ in Model II) and all the other particle cascade decays.
The red  line denotes the contribution from $\mu$ (or $e$), while the dashed
line denotes the contributions from the other particles. The left figure is 
for Model II with $m_D=1.8$ TeV, and the right figure  
for Model I with  $m_D=3.0$ TeV.}
\end{figure}

\begin{figure}[htb]
\begin{center}
\scalebox{1.1}{\input{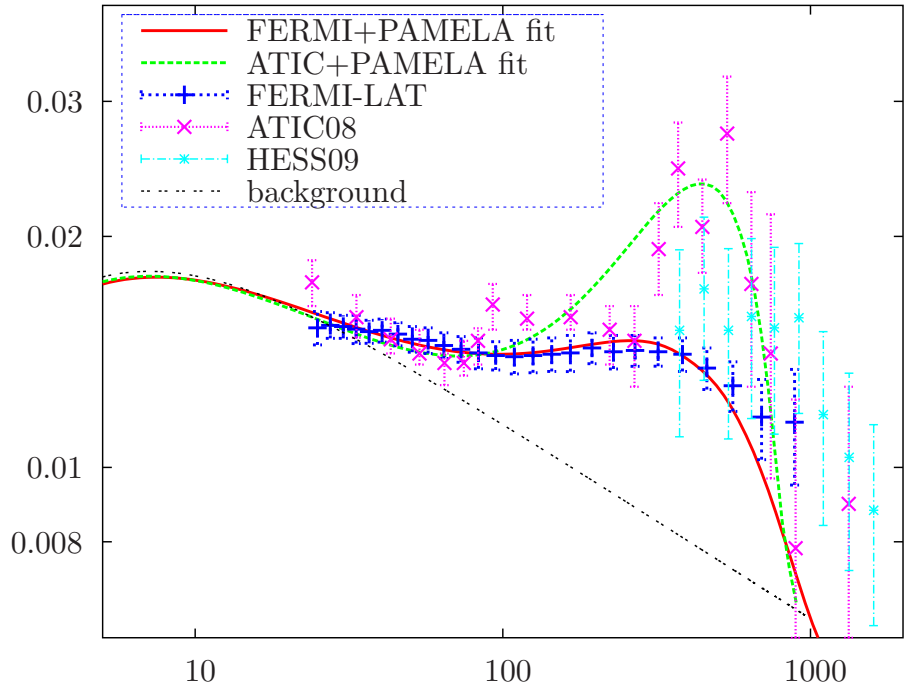}}
\end{center}
\caption{\label{FERMIATIC}  
The FERMI data fitting in Model I and the ATIC data
fitting in Model II. }
\end{figure}

The PAMELA positron fraction data and 
FERMI electron/positron data can be fitted very well by
choosing proper $\tau_{eff}$ and $\kappa$. The primary DM $S$ three-body
decays into  muons, which subsequently decay to electron/positron, give the
hard positron energy spectrum that is necessary to explain the steep
rise in positron fraction observed by the PAMELA experiment~\cite{Ibarra:2009dr}.
As for the FERMI experiment, its data show a rather flat spectrum
 up to about 1 TeV. Moreover, there are
a faint minimum and a faint peak  
at about 100 GeV and 400 GeV, respectively,
which can be considered as  the subtle fine structure.
Interestingly, we can fit such structure as well. 
The possible reason is  that: at low energy $(\lesssim 50$
GeV), the DM $S$ decay contributions to the observed electron/positron
spectrum are dominated by these from $A^0,H^0...$ channels; while at
high energy, they are dominated by these from cascade muon decays.
The muon decays produce a relative harder spectrum, 
which is peaked around 400 GeV. Because the
peak varies with the DM particle mass, this  explains why we choose  $m_D=3$
TeV as a typical fitting value in Model I. In Fig.~\ref{compare}, we compare 
these two contributions in Model I with $m_D=3.0$ TeV and 
in Model II with  $m_D=1.8$ TeV. 
In Fig.~\ref{FERMIATIC}, we present the fitting
for the FERMI data in Model I with  $\tau_{eff}=0.72\times 10^{26}s$, 
 $m_D=3.0$ TeV, and $\kappa=0.65$, and the fitting for the 
 ATIC data in Model II with $\tau_{eff}=0.52\times
10^{26}s$, $m_D=1.8$ TeV and $\kappa=0.67$. 
In short, these fittings are pretty good especially at 
relative low energy.

\begin{figure}[htb]
\begin{center}
\scalebox{1.0}{\input{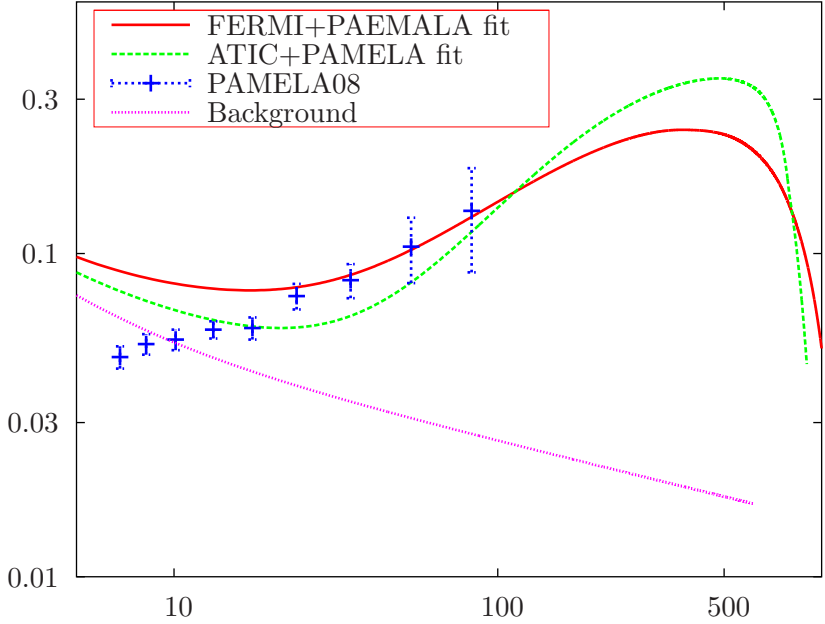}}
\end{center}
\caption{\label{pamela}  PAMELA  data fitting for positron fraction
 in Model I and Model II.}
\end{figure}

In addition,  from  Fig.~\ref{pamela}, we obtain that 
our fitting for the PAMELA positron fraction data in  Model II looks fine
since the positron spectrum from the DM direct three-body 
decay is relatively harder. However,
we can not fit  the PAMELA positron fraction data in  Model I
very well in the low energy region ($<20$ GeV). The reason is that the
low energy positron data are dominated by the sparticle and 
Higgs boson cascade decays, which produce large number 
of soft electron/positron. Therefore, the
positron fraction can exceed the background value. Interestingly, the main
character at the high energy region is reproduced pretty well. Especially, the
steep rising is obvious due to the harder positrons from muon decay.
Moreover, it predicts a continuously going up trend until arriving at the
peak around 0.5~TeV, which can be tested by the upcoming
 PAMELA experiment.

\subsection{Anti-Proton Fluxes}

Bcause Higgs  bosons and charginos can couple  to the SM quarks, 
their decays may produce anti-proton excess  as well. However, the
anti-proton excess is not observed by the PAMELA experiment. 
The propagation of anti-proton  is similar to that of positron, but it is
simpler by the virtue of no energy loss process from the inverse Compton
scattering
\begin{align}\label{pbar}
{\partial \psi\over \partial t}-\nabla\cdot({K(\vec{x},E)\nabla
\psi})+\nabla\cdot(\vec{V}_c(\vec{x})\psi)
=q(\vec{x},E)-2h\delta(z)\Gamma_{ann}\psi~.~\,
\end{align}
However, the Milky Way's galactic wind $\vec{V}_c(\vec{x})$, which is usually
assumed along the $z$ axial direction and then is reduced to $V_c\,{\rm
sign}(z)\vec{k}$, induces the drift of anti-protons during
propagation, as reflected in the third term. The last term
represents the annihilations between 
the anti-protons and interstellar protons in
the galactic plane, whose thickness is about $h~\approx ~0.1$~kpc~$\ll L$,
the half-thickness of cylinder diffusion region of charged
particles. The annihilation rate is given by
$\Gamma_{ann}=(n_H+4^{2/3}n_{He})\sigma_{p\bar p}^{ann}v_{\bar p}$
with~\cite{Maurin:2001sj}
\begin{align}
\sigma_{p\bar p}^{ann}=\{ \begin{array}{cc}
661(1+0.0115T^{-0.774}-0.984T^{0.0151}){\rm mbarn},\quad {\rm for}
\,\, T<15.5 \,{\rm GeV} \\ 36T^{-0.5}{\rm
mbarn},\quad\quad\quad\quad\quad\quad\quad\quad\quad\quad\quad\quad\quad\quad
{\rm for}\,\, T\geq15.5{\rm GeV}
\end{array}~,~\,
\end{align}
where $T=E-m_p$ is the proton kinetic energy. We can use 
$K_0$, $\delta$, $V_c$ and $L$ to classify the astrophysical
models that are compatible with the $B/C$ ratio into three
types~\cite{Lavalle:1900wn}, where $K_0$ and $\delta$ appear 
in the diffusion coefficient $K$. And each type is characterized 
by the flux of anti-protons. To
suppress the anti-proton flux, we choose the parameters 
which can produce the minimal anti-protons: $\delta=0.85$, $K_0=0.0016$
kpc$^2$Myr$^{-1}$, $L$=1 kpc, and $V_c=13.5$ km s$^{-1}$.

Similarly, the anti-proton flux at the heliosphere boundary is
formally solved to be
\begin{align}\label{pbarsolu}
\Phi_\odot^{\bar p}(E)=&{1\over \Gamma_{total}}\times{\beta_{\bar
p}\over 4\pi}\times{dN_{\bar p}^S\over dE}\times \mathcal {H}_{\bar
p}(E)~,~\,
\end{align}
where
\begin{align}
 {dN_{\bar p}^S\over
dx}
 =&{ \Gamma_{eff}\over \Gamma_{total}} \sum_{I,F}
\left(\mathcal {C}_{I,F}^2R_{I,F}\right)\times{1\over
R_{I,F}}
 {d\wt{N}\over dx_{I,F}}\otimes {dN\over dx^{I,F}_p }~.~
\end{align}
For details, please see Appendix B.

It is much simpler to obtain  the flux  for anti-proton than that of 
positron. The point is that the astrophysical information is totally 
factorized out in $\mathcal {H}_{\bar p}(E)$, and no further convolution
operation is needed. The predicted anti-proton fluxes 
comparing to the 
background flux in Model I and Model II are plotted in
Fig.~\ref{pbar}, where the suitable astrophysical parameters
have been chosen. The  background anti-protons mainly  arise from
the collisions between the primary CR protons (produced also by
supernova) and the interstellar hydrogen gas (for details, please see
Ref.~\cite{Maurin:2001sj}). 
Taking proper astrophysics parameters in the practical
calculations, we find that the anti-proton excess is definitely excluded 
below 50~GeV (Sixteen of the PAMELA seventeen points 
on $\bar p/p$ are in this energy region~\cite{Adriani:2008zr}.). 
In Model I for the
PAMELA and FERMI data fitting, the DM particle $S$ is very heavy about 
3~TeV, and then the  anti-proton flux is suppressed in all energy region
below about 130~GeV.  However, in Model II for the
PAMELA and ATIC data fitting, the  DM particle $S$ is about
$1.8$~TeV, and then the anti-proton flux begins to exceed background above
about 60 GeV. Especially, the excess is more significant as the
energy increasing, which is interesting since 
it can be tested in the future.

\begin{figure}[htb]
\begin{center}
\scalebox{1.0}{\input{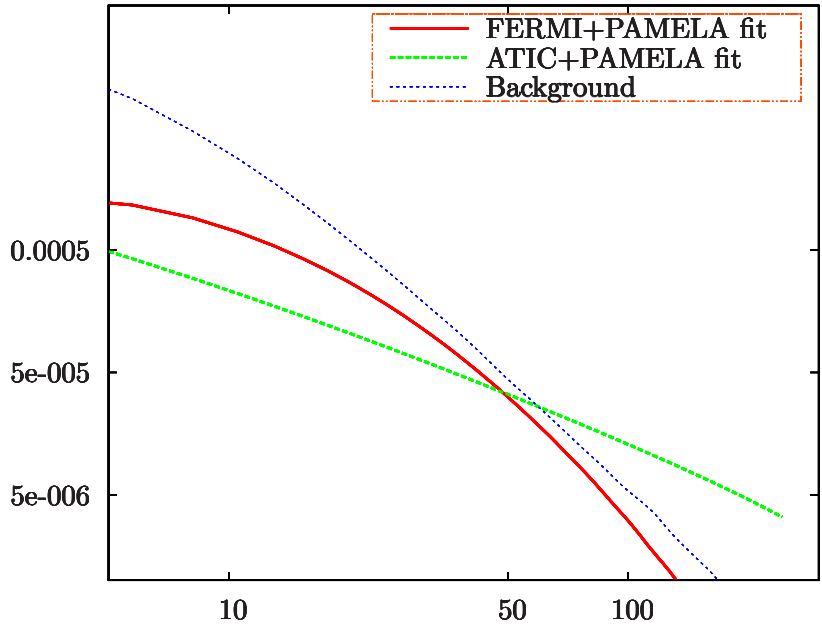}}
\end{center}
\caption{\label{pbar} PAMELA  data fitting for ${\bar p}/p$ ratio
in Model I and Model II. The  anti-proton fluxes
$\Phi_\odot^{\bar p}(E)$ is in the unit
GeV$^{-1}$m$^{-2}$s$^{-1}$sr$^{-1}$. }
\end{figure}

\section{Discussion and Conclusions \label{dis}}

We proposed two supersymmetric Standard Models with
decaying and stable DM particles. 
To explain the SM fermion
masses and mixings, we considered an anomalous $U(1)_X$ gauge 
symmetry whose anomaly is cancelled by the Green-Schwarz
mechanism. Around the string scale, the $U(1)_X$ gauge 
symmetry is broken down to the $Z_2$ symmetry under which
only $S$ is odd. Thus, $S$ is stable and can be a DM
candidate. After $S$ obtains a VEV around the TeV scale, 
the $Z_2$ symmetry is broken and then $S$
can decay. In our models,  on the one hand, the 
LSP neutralino has mass 101.6 GeV, and its relic
density is about $\Omega_{{\wt N}_1^0}h^2\approx 0.08$. 
Because the lightest neutralino-nucleon cross section is
$5\times 10^{-9}$ pb, the CDMS II results can be explained. 
On the other hand,
 $S$ is a decay DM particle that can three-body cascade 
decay into the MSSM particles.  With suitable $U(1)_X$ charges,
$S$ decays dominated into 
the second family of the SM leptons in Model I and into
the first family of the SM leptons in Model II. In Model I,
if the mass of the DM particle $S$ is 3 TeV
and $S$ has effective lifetime about $\tau_{eff}\approx0.72\times
10^{26}s$ (the real lifetime is about $2.2\times 10^{26}s$),
 we are able to explain the PAMELA/FERMI experiments
simultaneously. 
In Model II, if the mass of the DM particle $S$ is 1.8 TeV
and $S$ has effective lifetime about $\tau_{eff}\approx0.52\times
10^{26}s$, we can explain the PAMELA/ATIC experiments as well.
In addition, taking the proper astrophysics parameters that produce
the minimal anti-protons,  
we have shown that our models are consistent with
${\bar p}/p$ measurement in the PAMELA experiment.

Finally, although the accompanied gamma-ray and neutrino fluxes are not
discussed here, they deserve further study. The primary
hard neutrinos, as well as the soft gammas and neutrinos during the sparticle
and Higgs fragmentations are produced. They may provide the spectrum
properties through the inverse Compton scattering~\cite{Meade:2009iu} that
can be tested by the ongoing FERMI experiment. In addition, the 
complete and systematic studies of the multi-body decays and
multi-contributions are interesting since we do need them
in the DM model building. Furthermore,  it is interesting to
study the CMSSM parameter space where the LSP neutralino relic density is
smaller than the observed total DM relic desity since we may have
multicomponent DM.

%%%%%%%%%%%%%%%%%%%%%%%%%%%%%%%%%%%%%%%%%%%%%%%%%%%%%%%%%%%%%%%%%%%%%%%%%%%%%%%%

%%%%%%%%%%%%%%%%%%%%%%%%%%%%%%%%%%%%%%%%%%%%%%%%%%%%%%%%%%%%%%%%%%%%%%%%%%%%%%%%

%%%%%%%%%%%%%%%%%%%%%%%%%%%%%%%%%%%%%%%%%%%%%%%%%%%%

\begin{acknowledgments}

We would like to thank Xiao-Jun Bi, Xian Gao, Wan-Lei Guo, Chun-Li Tong,
 Peng-fei Yin, Qiang Yuan, and Xinmin Zhang for helpful discussions.
This research was supported in part 
by the Natural Science Foundation of China under grant No. 10821504,
by the DOE grant DE-FG03-95-Er-40917 (TL), and by
the Mitchell-Heep Chair in High Energy Physics (TL).

\end{acknowledgments}

\appendix

%%%%%%%%%%%%%%%%%%%%%%%%%%%%%%%%%%%%%%%%%%%%%%%%%%%%%%%%%%%%%%%%%%%%%%%%%%%%%%%%

%%%%%%%%%%%%%%%%%%%%%%%%%%%%%%%%%%%%%%%%%%%%%%%%%%%%%%%%%%%%%%%%%%%%%%%%%%%%%%%%

\section{The DM $S$ Three-body Decay}

In this Appendix, we develop some useful approximation formulae
for the three-body DM $S$ decay. First, let us
explain the convention. The decay width is calculated
in the rest frame of the heavy parent particle $S$ (Fermionic or
scalar component depends on the choice.). In a decay channel, we always have
the order of  the final state masses  as $m_3>m_2>m_1$.

In this case, in general $m_3$ is the mass of a slepton or Higgs
bonson, with $m_3/m_D\lesssim 0.1$. And $m_2$ denotes the mass of
Higgsino, $m_2/m_D\lesssim 0.05$. Finally, the lightest particle is
either  muon or neutrino, whose mass $m_1=m_{\mu,\nu} < 10^{-4}
m_D$ is ignorable. To simplify the calculations we set $m_1=0$.

\begin{figure}[htb]
\begin{center}
\includegraphics[width=3.5in]{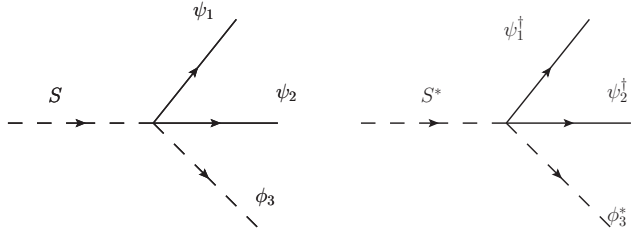}
\end{center}
\caption{\label{decay} The DM $S$
three-body decays into the Higgs bosons and leptons, or the
 sleptons, Higgsinos  as well as leptons.}
\end{figure}

First, we consider a general three-body decay  process described by  a
heavy complex scalar $S$ with mass $M$ that decays into two fermions
$\psi_{1,2}$ plus a scalar $\phi_3$.
 The Feynman diagram is given in Fig.~\ref{decay}.
It is not difficult to get  the corresponding amplitude square with the
final state spin summed over
\begin{align}\label{amplitude}
|{i\cal{M}}|^2= &4{\cal{C}}^2{\cal{C}}_I^2(p_1\cdot p_2+m_1m_2)~,~
\end{align}
where $\cal {C}$ is the coupling constant with mass dimension
$-1$, and ${\cal{C}}_I$ is the
possible mixing factor arising  from  converting the interaction state
into the mass eigenstate. The mass $m_1$ is still kept in the
momentum, and then will be set to zero finally.

\section{Fragmentation Function}

Fragmentation function $dN/dx_I$ is analytically calculable for
three-body decay $Y\rightarrow a+b+c$. For the isotropic decay it is
very convenient~\cite{bp} to write the phase space integral according
to the scaled variables $x_I=2P_I\cdot Y/\sqrt{Y^2}$,
%($\epsilon_a=m_a/Y$, x_a+a_b+x_c=2)
where the invariant mass is defined as $Y^2=(a+b+c)^2$. Here, 
 the particles and their four-momenta are labeled with the same symbols. 
In the rest frame of $Y$, we have $Y=m_Y$ and $x_I=2E_I/m_Y$. 
The three-body phase space is reduced to 
the two-body phase space by pairing $a$ with $b$ and
further integrating over $c$
\begin{align}
d_3(PS,Y\rightarrow abc)={Y^2\over 128\pi^3}  dx_adx_b~.
\end{align}
Then we find that  the spectrum of the observed final
state, for instance $a$, is obtained by integrating over
$x_b$. Kinematics constrains $x_a$ to be in the region
\begin{align}\label{limit}
2\epsilon_a\leqslant x_a\leqslant
1+\left(\epsilon_a^2-\epsilon_b^2-\epsilon_c^2-2\epsilon_b\epsilon_c\right)=R^{max}_a,
\end{align}
where $\epsilon_a^2\equiv m_a^2/Y^2$. If all the final state masses are
ignorable, the region is between 0 and 1. For fixed $x_a$,  $x_b$ is
constrained within the region $[R_b^{-},R_b^{+}]$
\begin{align}
R_b^{\pm}=& {1\over 2}(1-x_a+\epsilon_a^2)^{-1}
[(2-x_a)(1+\epsilon_a^2+\epsilon_b^2-\epsilon_c^2-x_a) \pm
 \sqrt{x_a^2-4\epsilon_a^2}\lambda^{1\over 2}
\left(1+\epsilon_a^2-x_a,\epsilon_b^2,\epsilon_c^2\right)]\cr
 \approx &{1\over 2}\left((2-x_a)\pm
 x_a\right),
\end{align}
where the approximation is valid for small mass limit
$\epsilon_{a,b,c}\ll 1$, and  the triangle function is defined as
$\lambda(x,y,z)=x^2+y^2+z^2-2xy-2xz-2yz$.

 The general process is $S\rightarrow \psi_1\psi_2\phi_3$
with amplitude given in Eq.~(\ref{amplitude}). 
In terms of $x$ and $\epsilon$, we can write the amplitude as follows
\begin{align}
|\mathcal {M}|^2=&2({\cal{C}{\cal{C}}_I})^2Y^2\left(1 + \epsilon_3^2
- (\epsilon_2+\epsilon_1)^2 -x_3\right)~,~\,
\end{align}
which is symmetric for the subscripts $1$ and $2$. The distribution
function of the fermion $\psi_{1}$ is obtained  by  paring $\psi_1$ and
$\phi_3$, {\it i.e.},  setting $a=\psi_1,\,b=\phi_3,\,c=\psi_2,\,$
%x_2=x_c is integrated over at first after using  the \delta function in pairing a and 3
\begin{align}{\label{SS}}
{dN\over dx_1}=&{{2({\cal{C}{\cal{C}}_I})^2Y^4 }\over
256\pi^3m_D}\times \mathcal {N} \int^{R_3^{+}(x_1)}_{R_3^-(x_1)}dx_3
\left(1 + \epsilon_3^2 - (\epsilon_2+\epsilon_1)^2 -x_3\right)\cr=&
{{2({\cal{C}{\cal{C}}_I})^2Y^4 }\over 256\pi^3m_D} \mathcal
{N}\left[1 + \epsilon_3^2 - (\epsilon_2+\epsilon_1)^2-{1\over
2}\left(R_3^+(x_1)+R_3^-(x_1)\right)\right]\cr&\times
\left(R_3^+(x_1)-R_3^-(x_1)\right),
\end{align}
where $\mathcal {N}=1/\Gamma$ is the normalization factor. In the
massless limit, it is approximated to be a simple function $x_1^2$.
So the fermion spectrum is very hard. The
distribution of $\psi_2$ is got simply by replacing the subscript 1
by 2 in the above formula. Similarly, we get the distribution
function of scalar
$\phi_3$ by choosing $a=\phi_3,\,b=\psi_1$, and $c=\psi_2$
%still by paring up $\phi_3$ and $\psi_1,c=2$,
\begin{align}
{dN\over dx_3}=& {{2({\cal{C}{\cal{C}}_I})^2Y^4 }\over
256\pi^3m_D}\mathcal {N} \int_{R_1^-(x_3)}^{R_1^+(x_3)}dx_1 \left(1
+ \epsilon_3^2 - (\epsilon_2+\epsilon_1)^2  -x_3\right)\cr
=&{{2({\cal{C}{\cal{C}}_I})^2Y^4 }\over 256\pi^3m_D} \mathcal
{N}\left(1 + \epsilon_3^2 - (\epsilon_2+\epsilon_1)^2
-x_3\right)\left(R_1^+(x_3)-R_1^-(x_3)\right).
\end{align}
Its massless limit is simplified to be ${dN/ dx_3}\rightarrow
(1-x_3)x_3$, so the scalar spectrum is softer than the  fermion 
spectrum. Integrating over $x_3$, we get the  decay rate 
%×îºóÒª³ËÒÔ{1/£¨2M)µÄÒò×Ó£¬Òª¼ÇµÃcheckÒ»ÏÂÕýÈ·µÄ³Ë×ÓÊÇ¶àÉÙ}
\begin{align}{\label{decayrate}}
\Gamma_I\approx& {1\over 128\pi^3}{1\over
m_D}\times({\cal{C}{\cal{C}}_I})^2Y^4\times R_{I,F} ~,~
\end{align}
where
\begin{align}
R_{I,F}=&\frac{1}{6}\left(1+9 \epsilon_3^2+12 \epsilon_3^2\log
\epsilon_3+12 \epsilon_2\epsilon_3-3\epsilon_2^2\right).
\end{align}
In this approximation, the terms involving $\epsilon_1$ is negligible
 here. Furthermore, if $\epsilon_2$ is also small
enough, the last two terms can be discarded.

For convenience, we define the DM effective decay  rate for
 all channels as the typical scale of its decay
\begin{align}{\label{decayrate}}
\Gamma_{eff}\equiv  {1\over 128\pi^3}{1\over
m_D}\times{\cal{C}}^2Y^4~.~
\end{align}
Then we have $\Gamma_{I}=\Gamma_{eff}{\cal{C}}_I^2R_{I,F}$. And the
normalized distribution  function is rewritten as a clear form
$dN/dx={1/R_{I,F}}\times (d\wt {N}/dx)$.

\subsection{General Cascade Decay Spectra}

In our models, $e^\pm$ signals come from  both the DM cascade decay
and direct decay, for example,
\begin{align}
S\rightarrow \wt N_{1,2} \mu \wt \mu,\quad  \mu \rightarrow e\nu
\nu, \quad\wt \mu\rightarrow \wt N_1 e.
\end{align}
The positron energy spectrum in rest frame of $S$ is determined to
be the convolution of two distribution functions. Physically
speaking,  the fragmentation functions
of the (N)MSSM particles $(I,F)$ can be extracted out from
 PYTHIA~\cite{Sjostrand:2006za}
\begin{align}
{dN\over dx^{I,F}_e},\quad x^{I,F}_e\equiv {2 E^{I,F}_e\over
m_{I,F}},
\end{align}
which is calculated in the rest frame of $(I, F)$ particle. Converting to the rest
frame of $S$, we obtain the total positron energy spectrum 
\begin{align}\label{convolution}
{dN\over dx^S_e}=& \sum_{I,F}{B_{I,F}}
\int^{R_{I,F}^{max}}_{2\epsilon_{I,F}} dx_{I,F}{1\over
R_{I,F}}{d\wt{N}\over dx_{I,F}}\int _{-1}^1 d \cos\theta_{I,F}
\int^{{\cal{I}}_e}_{2\epsilon_e} dx_e^I {dN\over dx^{I,F}_e}\cr &
\times \delta \left(2x_e^Sx_{I,F}^{-1} -x_e^{I,F}-\cos\theta_{I,F}
\sqrt{(x_e^{I,F})^2-4\epsilon_e^2} \sqrt{1-4\epsilon_{I,F}^2}\right)
,
\end{align}
with the scaled energies $ x^S_e\equiv 2 E^S_e/ m_D$ and $ x_{I,F}\equiv  2
E_{I,F}/ m_D$, and the mass ratios $\epsilon_e\equiv m_e/m_I$ and
$\epsilon_{I,F}\equiv m_{I,F}/m_D$. The branch ratio $B_{I,F}$ is
given by $B_{I,F}={\Gamma_{I,F}/\Gamma_{total}}$. 
%Because $\Gamma_{I,F}$ or  $B_{I,F}$  is only
%dependent on the $I$ but not $F$, so $F$ will be omitted from now on.
Notice that $\mathcal {I}_e$ denotes the largest positron energy
fraction in the fragmentation of the $({I,F})$ particle. 
But the positron production 
 is not clear in the course of fragmentation, so we cannot
 make sure the value of $\mathcal {I}_e$. However,
for $m_e\ll m_{I,F}$, we have $\mathcal {I}_e=1$~\footnote{If 
$({I,F})=\mu$, the positron  energy
fraction from the muon three-body
decay indeed has the upper limit about 1 from Eq. (\ref{limit}). However,
the positrons from the slepton or Higgs boson fragmentations tend
 to have a much softer spectrum. The PYTHIA simulation confirms this
point.}.

For three-body decays, the upper limit $R_{I,F}^{max}$  can be found
in Eq.~(\ref{limit}). In the rest frame of 
the $({I,F})$ particle, $\theta_{I,F}$ is
the angle between the electron spatial momentum direction and the boost axis
of the $({I,F})$ particle. 
The $\delta-$function simply indicates that for some fixed
boosted energy $x_e^S$, one should integrate over  all possible
configurations for $(x_e^{I,F},\cos\theta_{I,F} )$. In other words,
Eq.~(\ref{convolution}) makes the Lorentze boost of the energy spectrum:
$d{ N} / dx^{I,F}_e\rightarrow d{\wt N}/ dx^S_e$, and then convoluted by
the distribution $dN/dx_{I,F}$ that denotes the differential
probability of $S$ decay to $(I, F)$ particle 
in the rest frame of $S$. Finally, we
get the total positron  energy spectrum.

For isotropic decay of the $({I,F})$ particle, the distribution function is
independent on angle, which can be integrated out 
\begin{align}\label{11}
{dN\over dx^S_e}~= & {\Gamma_{eff}\over
\Gamma_{total}}\sum_{I,F}{\mathcal {C}_{I,F}^2R_{I,F}\over
\sqrt{1-4\epsilon_{I,F}^2}} \int^{R_{I,F}^{max}}_{2\epsilon_{I,F}}
dx_{I,F} {1\over R_{I,F}}{d\wt{N}\over dx_{I,F}}
\int^{max({x_{I,F}})}_{min({x_{I,F}})} dx_e^{I,F} \cr &
{1\over
\sqrt{(x_e^{I,F})^2-4\epsilon_e^2} }{dN\over dx^{I,F}_e}.
\end{align}
The normalization condition of $dN/dx^S_e$ in each
decay channel is
\begin{align}
\int {dN\over dx^S_e}d x_e^S=multi_{I,F}~,~\,
\end{align}
where $multi_{I,F}$ denotes how many electrons are produced by
 the fragmentation of each $(I,F)$ particle. 
The integrand region in Eq.~(\ref{11}) is given by
$\min(x_{I,F})=\mathcal {R}_-(x_{I,F})$ and
$\max(x_{I,F})=\min(\mathcal {{I}}_e,\mathcal {R}_+)$ with
\begin{align}
\mathcal {R}_\pm{(x_{I,F})}={1\over 2}\left( x_e^S \pm \sqrt{1
-4\epsilon_{I,F}^2}\sqrt{ (x_e^S)^2 -
    4(\epsilon_e \epsilon_{I,F} x_{I,F})^2}\right) \epsilon_{I,F}^{-2}
    x_{I,F}^{-1}.
\end{align}
If $\mathcal {R}_->\mathcal {I}_e$,  the integral is zero. Then the
constrained  $\mathcal{R}_\pm$, which depends only on the ratio
$r=x_{e}^S/x_{I,F}$, in turn constrains  the integral region for
$x_{I,F}$.  If $\epsilon_{I,F}\ll 1$ is satisfied, then $R_+\approx
x_e^S/(x_I \epsilon_I^2)\gg 1$. So the upper limit always takes
$\mathcal {{I}}_e$. For example, let us choose $(I,F)=\mu$. But in this
paper, this  is not always true since the sparticles and Higgs
bosons are heavy. Consequently there are some values of $r$ which gives
$\mathcal {R}_+>\mathcal {I}_e$. For the later case, one can
show that $\mathcal{R}_\pm$ is monotonically increase with $r$,
and the lower limit smaller than $\mathcal {I}_e$ gives the upper
bound for $r\leq\eta_2$, which  is the larger root of $\mathcal
{R}_-=0$. However, in the former case, $\mathcal {R}_->\mathcal
{I}_e$ gives bound $\eta_1\leq r\leq \eta_2$. In both cases, for fixed
$x_e^S$, we must have $x_I\geq x_e^S/\eta_2$.

\subsection{Analytical  $e^\pm$ Spectra from Three-Body Cascade Decays}

As an application, we use the above formulae to calculate the
$e^{\pm}$ spectrum functions from $S\rightarrow \mu+fermion+scalar$
followed by $\mu$ three-body decay. Notice that
$\epsilon_{\mu}\ll\epsilon_e\ll 1$ and $\eta_1\leq r\leq \eta_2$,
 we obtain $2\eta_1\epsilon_\mu\leq x_e^S \leq R_\mu^{max}
\eta_2$. The spectrum functions $dN/dx_e^{\mu}$   and $d\wt
N/dx_\mu$ are respectively given by 
\begin{align}\label{muon0}
{dN\over dx_e^\mu}= 2(x_e^\mu)^2(3-2x_e^\mu), \quad\quad  {d\wt
N\over dx_\mu}\approx \left({1\over 2}-{\epsilon_I^2\over
1-x_\mu}\right)x_\mu^2~,~
\end{align}
where the approximated spectrum is calculated according to
Eq.~(\ref{SS}). 

%In the modes $I$ denoted by $scalar$,  the mass
%of $fermion$ is even lighter. Thus, we only keep the leading order
%contribution from $scalar$.

 If we are only interested in the positrons with energy higher than
1 GeV, we obtain $\cal{R}_-$$\approx x_e^S/x_\mu$. Integrating over
$x_e^\mu$ and $x_\mu$, we have
\begin{align}\label{muon}
{dN\over dx^S_e}&=\int^{R_{\mu}^{max}}_{x_e^S/{\eta_2}}
dx_{\mu}\left[{5\over 3} - 3 \left({x_e^S\over x_\mu}\right)^2 +
{4\over3} \left({x_e^S\over x_\mu}\right)^3\right]\times
\left({1\over 2}-{\epsilon_I^2\over 1-x_\mu}\right)x_\mu^2\cr
&\approx {5\over 18}-{3\over 2}\left(x_e^S\right)^2+{11\over
9}\left(x_e^S\right)^3-{2\over 3}\left(x_e^S\right)^3\log x_e^S\cr
&+{\epsilon_I^2\over 3}\left[
5-5x_e^S+2\left(x_e^S\right)^2-2\left(x_e^S\right)^3+4\left(x_e^S\right)^3\log
x_e^S  \right. \cr & \left. 
+2\left(5-9\left(x_e^S\right)^2+4\left(x_e^S\right)^3\right)
\right].\cr
\end{align}
The correction term is complicated but important since without it we 
cannot have the right order. This expression is valid for the
energy fraction $2\epsilon_\mu\eta_2\leq x_e^S\leq\eta_2 R^{max}_\mu
$. The DM three-body cascade decay produces a harder spectrum than the
two-body decay since its intermediate particle produces a
hard spectrum. This is clearly shown in the muon case.

\section {Fitting  Positron  Spectra}

In this Appendix, we present the positron spectra got through the above
procedure. Similarly, we can calculate the anti-proton spectra,
although we will not present the details here.
The basic method is to simulate the positron and anti-proton
spectra in the rest frame of the sparticles and Higgs bosons
via PYTHIA (each state ($I,F$) must be simulated separately 
and then summed over). We will not pay much attention to the
detail fitting but present the results directly. Notation is
consistent with the one given above, but we change the variable to
energy, as in the original form from simulation.

The spectra are  divided into three types: the type of sleptons, the type of
Higgs bosons and the type of Higgsinos.  For sleptons, the spectra from
the left- and right-handed sparticles have quite different behaviours, since
 their weak boson decays are different. The total
$e^\pm$ from fragmentations are described by the following functions 
\begin{align}
\wt \mu_L: \quad{dN\over dE}&= \exp^{16.926-15.601 E^{0.183}}
     E^{1.438} \left[1 + 14.90 \log(1 + 0.05 E)^{6.1}\right]-8.5\times10^{-5}
     E^{0.55},\cr
{\wt {\nu_\mu}}_L: \quad{dN\over dE}&= \exp^{16.913-15.601
E^{0.183}}
     E^{1.437} \left[1 + 14.9 \log(1 + 0.055 E)^{6.5}\right]-9.0\times10^{-7}
     E^{0.60},\cr
     \wt {\mu}_R: \quad{dN\over dE}&= {E^{1.302}\over
 0.245+
  262.353 E^{1.971}} + {E^{0.031}\over 104.334+ 5.757\times 10^{-5}E^{3.242}}~.~
\end{align}
The type for all the Higgs bosons have the similar form,
probably because they all are dominated by the $H_d$ components.
For simplicity, we only present $A_0$ as follows
\begin{align}
     A^0: \quad{dN\over dE}= \exp^{17.716-15.601 E^{0.160}}
     E^{1.327} \left[1 + 4.0 \log(1 + 0.021
E)^{3.0}\right]-1.0\times10^{-6}E^{0.8}.
     \end{align}
At last, the fitting functions for the neutralinos $\wt C^0_{3,4}$ and
charginos $\wt C_{1,2}^{\pm}$ are given by     
\begin{align}
{\wt {C}}_3^0: \quad{dN\over dE}= \exp^{15.461-15.601 E^{0.163}}
     E^{1.311} \left[1 + 40.0 \log(1 + 0.025 E)^{2.4}\right]-2.7\times10^{-5}
     E^{0.7},\cr
{\wt {C}}_4^0: \quad{dN\over dE}= \exp^{17.880-15.601 E^{0.174}}
     E^{1.408} \left[1 + 14.0 \log(1 + 0.024 E)^{4.2}\right]-1.7\times10^{-6}
     E^{0.5},\cr
{\wt {C}}_1^{\pm}: \quad{dN\over dE}= \exp^{17.198-15.601 E^{0.195}}
     E^{1.494} \left[1 - 1.1 \log(1 + 0.045 E)^{3.0}\right]-2.7\times10^{-6}
     E^{0.5},\cr
{\wt {C}}_2^{\pm}: \quad{dN\over dE}= \exp^{17.814-15.601 E^{0.174}}
     E^{1.416} \left[1 + 13.0 \log(1 + 0.026 E)^{3.5}\right]+1.4\times10^{-6}
     E^{1.4}~.~\,
\end{align}
\begin{figure}[htb]
\begin{center}
\includegraphics[width=6.0in]{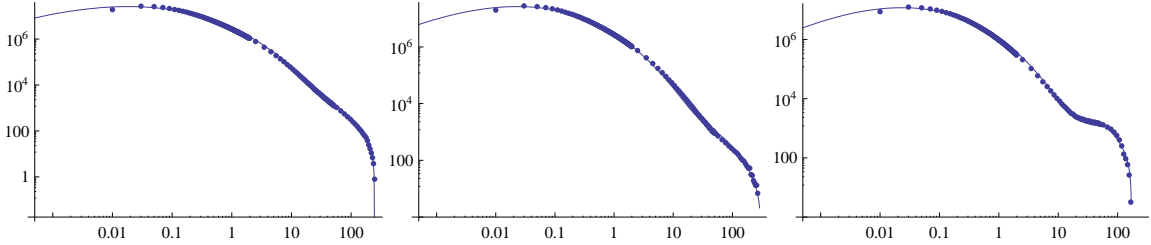}
\end{center}
\caption{\label{fitting} Typical  fitting functions: $\wt C_4^0$
(left), $H$ (middle) and $\wt \mu_L$ (right). In the simulations,
$3\times 10^5$ events have been generated by  PYTHIA.}
\end{figure}
 Basically, these fitting functions consist of
three parts.  The first part describes the low energy region very
well, the second logarithmic part is added to slow down the exceptional
decreasing,  and the last part is used to modulate the region near
the DM mass cutoff region.  This kind of fitting functions is not
universal, and only applies to our parameter sets. We present
the typical fittings for three types in Fig.~\ref{fitting}.
By the way, the small 
discrepancies between the fitting functions and simulations 
do exist at high energy. So the fitting functions are simply used as 
the referred fitting functions.

%%%%%%%%%%%%%%%%%%%%%%%%%%%%%%%%%%%%%%%%%%%%%%%%%%%%%%%%%%%%%%%%%%%%%%%%%%%%%%%%

%%%%%%%%%%%%%%%%%%%%%%%%%%%%%%%%%%%%%%%%%%%%%%%%%%%%%%%%%%%%%%%%%%%%%%%%%%%%%%%%


\begin{thebibliography}{99}
\itemsep 0.5mm

%%%%%%%%%%%%%%%%%%%%%%%%%%%%%%%%%%%%%%%%%%%%%%%%%%%%%%%%%%%%%%%%%%%%%%

%%%%%%%%%%%%%%%%%%%%%%%%%%%%%%%%%%%%%%%%%%%%%%%%%%%%%%%%%%%%%%%%%%%%%%

%%%%%%%%%%%%%%%%%%%%%%%%%%%%%%%%%%%%%%%%%%%%%%%%%%%%%%%%%%%%%%%%%%%%%%

%%%%%%%%%%%%%%%%%%%%%%%%%%%%%%%%%%%%%%%%%%%%%%%%%%%%%%%%%%%%%%%%%%%%%%



%%%%%%%%%%%%%%%%%%%%%%%%%%%%%%%%%%%%%%%%%%%%%%%%%%%%%%%%%%%%%%%%%%%%%%

%%%%%%%%%%%%%%%%%%%%%%%%%%%%%%%%%%%%%%%%%%%%%%%%%%%%%%%%%%%%%%%%%%%%%%


\bibitem{Chang:2008zz}
  J.~Chang {\it et al.},
%  ``An Excess of Cosmic Ray Electrons at Energies Of 300-800 GeV,''
  Nature {\bf 456}, 362 (2008).
  %%CITATION = NATUA,456,362;%%

\bibitem{Torii:2008}
  S.~Torii {\it et al.} [PPB-BETS Collaboration],
%  ``High-energy electron observations by PPB-BETS flight in Antarctica,''
  arXiv:0809.0760 [astro-ph].
  %%CITATION = ARXIV:0809.0760;%%

%\cite{Adriani:2008zr}
\bibitem{Adriani:2008zr}
  O.~Adriani {\it et al.}  [PAMELA Collaboration],
%  ``An anomalous positron abundance in cosmic rays with energies 1.5-100 GeV,''
  Nature {\bf 458}, 607 (2009).
%  [arXiv:0810.4995 [astro-ph]].
  %%CITATION = NATUA,458,607;%%

%\cite{Adriani:2008zq}
\bibitem{Adriani:2008zq}
  O.~Adriani {\it et al.},
%  ``A new measurement of the antiproton-to-proton flux ratio up to 100 GeV in
%  the cosmic radiation,''
  Phys.\ Rev.\ Lett.\  {\bf 102}, 051101 (2009).
%  [arXiv:0810.4994 [astro-ph]].
  %%CITATION = PRLTA,102,051101;%%


%\cite{Abdo:2009zk}
\bibitem{Abdo:2009zk}
  A.~A.~Abdo {\it et al.}  [The Fermi LAT Collaboration],
  %``Measurement of the Cosmic Ray e+ plus e- spectrum from 20 GeV to 1 TeV with
  %the Fermi Large Area Telescope,''
  Phys.\ Rev.\ Lett.\  {\bf 102}, 181101 (2009).
%  [arXiv:0905.0025 [astro-ph.HE]].
  %%CITATION = PRLTA,102,181101;%%


%\cite{Aharonian:2009ah}
\bibitem{Aharonian:2009ah}
  H.~E.~S.~S.~Collaboration,
%  ``Probing the ATIC peak in the cosmic-ray electron spectrum with H.E.S.S,''
  arXiv:0905.0105 [astro-ph.HE].
  %%CITATION = ARXIV:0905.0105;%%

\bibitem{HESS:2008aa}
  F.~Aharonian {\it et al.}  [H.E.S.S. Collaboration],
%  ``The energy spectrum of cosmic-ray electrons at TeV energies,''
  Phys.\ Rev.\ Lett.\  {\bf 101}, 261104 (2008).
%  [arXiv:0811.3894 [astro-ph]].
  %%CITATION = PRLTA,101,261104;%%


%%%%%%%%%%%%%%%%%%%%%%%%%%%%%%%%%%%%%%%%%%%%%%%%%%%%%%%%%%%%%%%%%%%%%%

%%%%%%%%%%%%%%%%%%%%%%%%%%%%%%%%%%%%%%%%%%%%%%%%%%%%%%%%%%%%%%%%%%%%%%




%%%%%%%%%%%%%%%%%%%%%%%%%%%%%%%%%%%%%%%%%%%%%%%%%%%%%%%%%%%%%%%%%%%%%%

%%%%%%%%%%%%%%%%%%%%%%%%%%%%%%%%%%%%%%%%%%%%%%%%%%%%%%%%%%%%%%%%%%%%%%



%\cite{Lavalle:1900wn}
\bibitem{Lavalle:1900wn}
  J.~Lavalle, Q.~Yuan, D.~Maurin and X.~J.~Bi,
  %``Full Calculation of Clumpiness Boost factors for Antimatter Cosmic Rays in
  %the light of \LambdaCDM N-body simulation results,''
  arXiv:0709.3634 [astro-ph].
  %%CITATION = ARXIV:0709.3634;%%



%\cite{Hisano:2003ec}
\bibitem{Hisano:2003ec}
  J.~Hisano, S.~Matsumoto and M.~M.~Nojiri,
  %``Explosive dark matter annihilation,''
  Phys.\ Rev.\ Lett.\  {\bf 92}, 031303 (2004);
% arXiv:hep-ph/0307216.
%\cite{Cirelli:2008id}
%\bibitem{Cirelli:2008id}
  M.~Cirelli, R.~Franceschini and A.~Strumia,
  %``Minimal Dark Matter predictions for galactic positrons, anti-protons,
  %photons,''
  Nucl.\ Phys.\  B {\bf 800}, 204 (2008).
%  arXiv:0802.3378 [hep-ph].
  %%CITATION = NUPHA,B800,204;%%


%\cite{ArkaniHamed:2008qn}
\bibitem{ArkaniHamed:2008qn}
  N.~Arkani-Hamed, D.~P.~Finkbeiner, T.~R.~Slatyer and N.~Weiner,
  %``A Theory of Dark Matter,''
  Phys.\ Rev.\  D {\bf 79}, 015014 (2009).
%  arXiv:0810.0713 [hep-ph].
  %%CITATION = PHRVA,D79,015014;%%

%\cite{Nomura:2008ru}
\bibitem{Nomura:2008ru}
  Y.~Nomura and J.~Thaler,
  %``Dark Matter through the Axion Portal,''
  Phys.\ Rev.\  D {\bf 79}, 075008 (2009).
%  arXiv:0810.5397 [hep-ph].
  %%CITATION = PHRVA,D79,075008;%%


%\cite{Ibe:2008ye}
\bibitem{Ibe:2008ye}
D.~Feldman, Z.~Liu and P.~Nath,
  %``PAMELA Positron Excess as a Signal from the Hidden Sector,''
  Phys.\ Rev.\  D {\bf 79}, 063509 (2009);
%  [arXiv:0810.5762 [hep-ph]].
  %%CITATION = PHRVA,D79,063509;%%
  M.~Ibe, H.~Murayama and T.~T.~Yanagida,
  %``Breit-Wigner Enhancement of Dark Matter Annihilation,''
  Phys.\ Rev.\  D {\bf 79}, 095009 (2009);
%arXiv:0812.0072 [hep-ph].
  %%CITATION = PHRVA,D79,095009;%%
%\cite{Guo:2009aj}
%\bibitem{Guo:2009aj}
  W.~L.~Guo and Y.~L.~Wu,
  %``Enhancement of Dark Matter Annihilation via Breit-Wigner Resonance,''
  Phys.\ Rev.\  D {\bf 79}, 055012 (2009);
%  arXiv:0901.1450 [hep-ph].
  %%CITATION = PHRVA,D79,055012;%%
%\cite{Bi:2009uj}
%\bibitem{Bi:2009uj}
  X.~J.~Bi, X.~G.~He and Q.~Yuan,
  %``Parameters in a class of leptophilic models from PAMELA, ATIC and FERMI,''
  Phys. Lett. {\bf B678}, 168 (2009).
%arXiv:0903.0122 [hep-ph].
  %%CITATION = ARXIV:0903.0122;%%

%\cite{Bi:2009am}
\bibitem{Bi:2009am}
  X.~J.~Bi, R.~Brandenberger, P.~Gondolo, T.~Li, Q.~Yuan and X.~m.~Zhang,
  %``Non-Thermal Production of WIMPs, Cosmic $e^\pm$ Excesses and $\gamma$-rays
  %from the Galactic Center,''
  Phys.\ Rev.\  D {\bf 80}, 103502 (2009).
%  [arXiv:0905.1253 [hep-ph]].
  %%CITATION = PHRVA,D80,103502;%%

%\cite{Chen:2008dh}
\bibitem{Chen:2008dh}
  C.~R.~Chen and F.~Takahashi,
  %``Cosmic rays from Leptonic Dark Matter,''
  JCAP {\bf 0902}, 004 (2009).
%  [arXiv:0810.4110 [hep-ph]].
  %%CITATION = JCAPA,0902,004;%%

%\cite{Yin:2008bs}
\bibitem{Yin:2008bs}
  P.~f.~Yin, Q.~Yuan, J.~Liu, J.~Zhang, X.~j.~Bi and S.~h.~Zhu,
  %``PAMELA data and leptonically decaying dark matter,''
  Phys.\ Rev.\  D {\bf 79}, 023512 (2009).
%  [arXiv:0811.0176 [hep-ph]].
  %%CITATION = PHRVA,D79,023512;%%


%\cite{Arvanitaki:2008hq}
\bibitem{Arvanitaki:2008hq}
  A.~Arvanitaki, S.~Dimopoulos, S.~Dubovsky, P.~W.~Graham, R.~Harnik and S.~Rajendran,
  %``Astrophysical Probes of Unification,''
  Phys.\ Rev.\  D {\bf 79}, 105022 (2009).
%  arXiv:0812.2075 [hep-ph].
  %%CITATION = PHRVA,D79,105022;%%


%\cite{Ibarra:2009dr}
\bibitem{Ibarra:2009dr}
  A.~Ibarra, D.~Tran and C.~Weniger,
  %``Decaying Dark Matter in Light of the PAMELA and Fermi LAT Data,''
  arXiv:0906.1571 [hep-ph].
  %%CITATION = ARXIV:0906.1571;%%

%decaying DM
%\cite{Nardi:2008ix}
\bibitem{Nardi:2008ix}
  E.~Nardi, F.~Sannino and A.~Strumia,
  %``Decaying Dark Matter can explain the electron/positron excesses,''
  JCAP {\bf 0901}, 043 (2009);
%  arXiv:0811.4153 [hep-ph].
  %%CITATION = JCAPA,0901,043;%%
%\cite{Ruderman:2009ta}
%\bibitem{Ruderman:2009ta}
  J.~T.~Ruderman and T.~Volansky,
  %``Searching for Smoking Gun Signatures of Decaying Dark Matter,''
  arXiv:0907.4373 [hep-ph];
  %%CITATION = ARXIV:0907.4373;%%
%\cite{Ibarra:2008jk}
%\bibitem{Ibarra:2008jk}
  A.~Ibarra and D.~Tran,
  %``Decaying Dark Matter and the PAMELA Anomaly,''
  JCAP {\bf 0902}, 021 (2009);
%  arXiv:0811.1555 [hep-ph].
  %%CITATION = JCAPA,0902,021;%%
%\cite{Luo:2009xd}
%\bibitem{Luo:2009xd}
  M.~Luo, L.~Wang, W.~Wu and G.~Zhu,
  %``Decaying Dark Matter in Supersymmetric SU(5) Models,''
  arXiv:0911.3235 [hep-ph].
  %%CITATION = ARXIV:0911.3235;%%

%\cite{Liu:2009sq}
\bibitem{Liu:2009sq}
  J.~Liu, Q.~Yuan, X.~Bi, H.~Li and X.~Zhang,
  %``A Markov Chain Monte Carlo Study on Dark Matter Property Related to the
  %Cosmic e$^{\pm}$ Excesses,''
  arXiv:0906.3858 [astro-ph.CO].
  %%CITATION = ARXIV:0906.3858;%%


%%%%%%%%%%%%%%%%%%%%%%%%%%%%%%%%%%%%%%%%%%%%%%%%%%%%%%%%%%%%%%%%%%
%%%%%%%%%%%%%%%%%%%%%%%%%%%%%%%%%%%%%%%%%%%%%%%%%%%%%%%%%%%%%%%%%%

%\cite{Ahmed:2009zw}
\bibitem{Ahmed:2009zw}
  Z.~Ahmed {\it et al.}  [The CDMS-II Collaboration],
  %``Results from the Final Exposure of the CDMS II Experiment,''
  arXiv:0912.3592 [astro-ph.CO].
  %%CITATION = ARXIV:0912.3592;%%




%\cite{Kadastik:2009ca}
\bibitem{Kadastik:2009ca}
M.~Kadastik, K.~Kannike, A.~Racioppi and M.~Raidal,
%``EWSB from the soft portal into Dark Matter and prediction for direct
%detection,''
arXiv:0912.2729 [hep-ph];
%%CITATION = ARXIV:0912.2729;%%
%\cite{Kadastik:2009gx}
%\bibitem{Kadastik:2009gx}
M.~Kadastik, K.~Kannike, A.~Racioppi and M.~Raidal,
%``Implications of the CDMS result on Dark Matter and LHC physics,''
arXiv:0912.3797 [hep-ph];
%%CITATION = ARXIV:0912.3797;%%
%\cite{Bernal:2009jc}
%\bibitem{Bernal:2009jc}
N.~Bernal and A.~Goudelis,
%``Dark matter detection in the BMSSM,''
arXiv:0912.3905 [hep-ph];
%%CITATION = ARXIV:0912.3905;%%
%\cite{Bottino:2009km}
%\bibitem{Bottino:2009km}
A.~Bottino, F.~Donato, N.~Fornengo and S.~Scopel,
%``Relic neutralinos and the two dark matter candidate events of theCDMS II
%experiment,''
arXiv:0912.4025 [hep-ph];
%%CITATION = ARXIV:0912.4025;%%
%\cite{Feldman:2009pn}
%\bibitem{Feldman:2009pn}
D.~Feldman, Z.~Liu and P.~Nath,
%``Connecting the Direct Detection of Dark Matter with Observation of
%Sparticles at the LHC,''
arXiv:0912.4217 [hep-ph];
%%CITATION = ARXIV:0912.4217;%%
%\cite{Kopp:2009qt}
%\bibitem{Kopp:2009qt}
J.~Kopp, T.~Schwetz and J.~Zupan,
%``Global interpretation of direct Dark Matter searches after CDMS-II
%results,''
arXiv:0912.4264 [hep-ph];
%%CITATION = ARXIV:0912.4264;%%
%\cite{Allahverdi:2009sb}
%\bibitem{Allahverdi:2009sb}
R.~Allahverdi, B.~Dutta and Y.~Santoso,
%``Models of supersymmetric dark matter and their predictions in light of
%CDMS,''
arXiv:0912.4329 [hep-ph];
%%CITATION = ARXIV:0912.4329;%%
%\cite{Endo:2009uj}
%\bibitem{Endo:2009uj}
M.~Endo, S.~Shirai and K.~Yonekura,
%``Phenomenological Aspects of Gauge Mediation with Sequestered
%Supersymmetry
%Breaking in light of Dark Matter Detection,''
arXiv:0912.4484 [hep-ph];
%%CITATION = ARXIV:0912.4484;%%
%\cite{Holmes:2009uu}
%\bibitem{Holmes:2009uu}
M.~Holmes and B.~D.~Nelson,
%``Non-Universal Gaugino Masses, CDMS, and the LHC,''
arXiv:0912.4507 [hep-ph];
%%CITATION = ARXIV:0912.4507;%%
%\cite{Cao:2009uw}
%\bibitem{Cao:2009uw}
Q.~H.~Cao, C.~R.~Chen, C.~S.~Li and H.~Zhang,
%``Effective Dark Matter Model: Relic density, CDMS II, Fermi LAT and
%LHC,''
arXiv:0912.4511 [hep-ph];
%%CITATION = ARXIV:0912.4511;%%
%\cite{Cheung:2009wb}
%\bibitem{Cheung:2009wb}
K.~Cheung and T.~C.~Yuan,
%``Implication on Higgs invisible width in light of the new CDMS result,''
arXiv:0912.4599 [hep-ph];
%%CITATION = ARXIV:0912.4599;%%
%\cite{Hisano:2009xv}
%\bibitem{Hisano:2009xv}
J.~Hisano, K.~Nakayama and M.~Yamanaka,
%``Implications of CDMS II result on Higgs sector in the MSSM,''
arXiv:0912.4701 [hep-ph];
%%CITATION = ARXIV:0912.4701;%%
%\cite{He:2009yd}
%\bibitem{He:2009yd}
X.~G.~He, T.~Li, X.~Q.~Li, J.~Tandean and H.~C.~Tsai,
%``The Simplest Dark-Matter Model, CDMS II Results, and Higgs Detection at
%LHC,''
arXiv:0912.4722 [hep-ph];
%%CITATION = ARXIV:0912.4722;%%
%\cite{Asano:2010yi}
%\bibitem{Asano:2010yi}
M.~Asano and R.~Kitano,
%``Constraints on Scalar Phantoms,''
arXiv:1001.0486 [hep-ph];
%%CITATION = ARXIV:1001.0486;%%
%\cite{Gogoladze:2009mc}
%\bibitem{Gogoladze:2009mc}
I.~Gogoladze, R.~Khalid, S.~Raza and Q.~Shafi,
%``CDMS II Inspired Neutralino Dark Matter in Flipped SU(5),''
arXiv:0912.5411 [hep-ph];
%%CITATION = ARXIV:0912.5411;%%
%\cite{Aoki:2009pf}
%\bibitem{Aoki:2009pf}
M.~Aoki, S.~Kanemura and O.~Seto,
%``Multi-Higgs portal dark matter under the CDMS-II results,''
arXiv:0912.5536 [hep-ph];
%%CITATION = ARXIV:0912.5536;%%
%\cite{Foot:2010th}
%\bibitem{Foot:2010th}
R.~Foot,
%``Relevance of the CDMSII events for mirror dark matter,''
arXiv:1001.0096 [hep-ph];
%%CITATION = ARXIV:1001.0096;%%
%\cite{Guo:2010vy}
%\bibitem{Guo:2010vy}
W.~L.~Guo, Y.~L.~Wu and Y.~F.~Zhou,
%``Exploration of decaying dark matter in a left-right symmetric model,''
arXiv:1001.0307 [hep-ph];
%%CITATION = ARXIV:1001.0307;%%
%\cite{Shu:2010ta}
%\bibitem{Shu:2010ta}
J.~Shu, P.~f.~Yin and S.~h.~Zhu,
%``Neutrino Constraints on Inelastic Dark Matter after CDMS II,''
arXiv:1001.1076 [hep-ph].
%%CITATION = ARXIV:1001.1076;%%


%%%%%%%%%%%%%%%%%%%%%%%%%%%%%%%%%%%%%%%%%%%%%%%%%%%%%%%%%%%%%%%%%%
%%%%%%%%%%%%%%%%%%%%%%%%%%%%%%%%%%%%%%%%%%%%%%%%%%%%%%%%%%%%%%%%%%



\bibitem{MC-DMs}
 C.~Boehm, P.~Fayet and J.~Silk,
%``Light and heavy dark matter particles,''
Phys.\ Rev.\ D {\bf 69}, 101302 (2004);
% [arXiv:hep-ph/0311143].
%%CITATION = PHRVA,D69,101302;%%
E.~Ma,
%``Supersymmetric Model of Radiative Seesaw Majorana Neutrino Masses,''
Annales Fond.\ Broglie {\bf 31}, 285 (2006);
% [arXiv:hep-ph/0607142].
%%CITATION = AFLBD,31,285;%
Q.~H.~Cao, E.~Ma, J.~Wudka and C.~P.~Yuan,
  %``Multipartite Dark Matter,''
  arXiv:0711.3881 [hep-ph];
  %%CITATION = ARXIV:0711.3881;%%
M.~Adibzadeh and P.~Q.~Hung,
%``The relic density of shadow dark matter candidates,''
Nucl.\ Phys.\ B {\bf 804}, 223 (2008);
% [arXiv:0801.4895 [astro-ph]].
J.~L.~Feng and J.~Kumar,
%``The Wimpless Miracle: Dark-Matter Particles Without Weak-Scale Masses Or
%Weak Interactions,''
Phys.\ Rev.\ Lett.\ {\bf 101}, 231301 (2008);
%[arXiv:0803.4196 [hep-ph]].
%%CITATION = PRLTA,101,231301;%%
H.~Sung Cheon, S.~K.~Kang and C.~S.~Kim,
%``Doubly Coexisting Dark Matter Candidates in an Extended Seesaw Model,''
Phys.\ Lett.\ B {\bf 675}, 203 (2009);
% [arXiv:0807.0981 [hep-ph]].
%%CITATION = PHLTA,B675,203;%%
%%CITATION = NUPHA,B804,223;%%
J.~H.~Huh, J.~E.~Kim and B.~Kyae,
%``Two dark matter components in N_{DM}MSSM and PAMELA data,''
Phys.\ Rev.\ D {\bf 79}, 063529 (2009);
% [arXiv:0809.2601 [hep-ph]].
%%CITATION = PHRVA,D79,063529;%%
M.~Fairbairn and J.~Zupan,
%``Two component dark matter,''
JCAP {\bf 0907}, 001 (2009),[Published Version];
%%CITATION = JCAPA,0907,001;%%
K.~M.~Zurek,
%``Multi-Component Dark Matter,''
Phys.\ Rev.\ D {\bf 79}, 115002 (2009);
%[arXiv:0811.4429 [hep-ph]].
%%CITATION = PHRVA,D79,115002;%%
B.~Batell, M.~Pospelov and A.~Ritz,
%``Direct Detection of Multi-component Secluded WIMPs,''
Phys.\ Rev.\ D {\bf 79}, 115019 (2009);
%[arXiv:0903.3396 [hep-ph]].
%%CITATION = PHRVA,D79,115019;%%
S.~Profumo, K.~Sigurdson and L.~Ubaldi,
%``Can we discover multi-component WIMP dark matter?,''
JCAP {\bf 0912}, 016 (2009);
%[arXiv:0907.4374 [hep-ph]].
%%CITATION = JCAPA,0912,016;%
 F.~Chen, J.~M.~Cline and A.~R.~Frey,
 %``Nonabelian dark matter: models and constraints,''
 Phys.\ Rev.\  D {\bf 80}, 083516 (2009);
% [arXiv:0907.4746 [hep-ph]].
 %%CITATION = PHRVA,D80,083516;%%
  H.~Zhang, C.~S.~Li, Q.~H.~Cao and Z.~Li,
  %``A Dark Matter Model with Non-Abelian Gauge Symmetry,''
  arXiv:0910.2831 [hep-ph].
  %%CITATION = ARXIV:0910.283%


%\cite{Hur:2007ur}
\bibitem{Hur:2007ur}
  T.~Hur, H.~S.~Lee and S.~Nasri,
  %``A Supersymmetric U(1) -prime model with multiple dark matters,''
  Phys.\ Rev.\  D {\bf 77}, 015008 (2008).
%  [arXiv:0710.2653 [hep-ph]].
  %%CITATION = PHRVA,D77,015008;%%


%\cite{Froggatt:1978nt}
\bibitem{Froggatt:1978nt}
  C.~D.~Froggatt and H.~B.~Nielsen,
  %``Hierarchy Of Quark Masses, Cabibbo Angles And CP Violation,''
  Nucl.\ Phys.\  B {\bf 147}, 277 (1979).
  %%CITATION = NUPHA,B147,277;%%

%\cite{Green:1984sg}
\bibitem{Green:1984sg}
  M.~B.~Green and J.~H.~Schwarz,
  %``Anomaly Cancellation In Supersymmetric D=10 Gauge Theory And Superstring
  %Theory,''
  Phys.\ Lett.\  B {\bf 149}, 117 (1984).
  %%CITATION = PHLTA,B149,117;%%


%\cite{Ibanez:1994ig}
\bibitem{Ibanez:1994ig}
  L.~E.~Ibanez and G.~G.~Ross,
  %``Fermion masses and mixing angles from gauge symmetries,''
  Phys.\ Lett.\  B {\bf 332}, 100 (1994);
%  [arXiv:hep-ph/9403338].
  %%CITATION = PHLTA,B332,100;%%
V.~Jain and R.~Shrock,
  %``Models of fermion mass matrices based on a flavor dependent and generation
  %dependent U(1) gauge symmetry,''
  Phys.\ Lett.\  B {\bf 352}, 83 (1995);
%  [arXiv:hep-ph/9412367].
  %%CITATION = PHLTA,B352,83;%%
E.~Dudas, S.~Pokorski and C.~A.~Savoy,
  %``Yukawa matrices from a spontaneously broken Abelian symmetry,''
  Phys.\ Lett.\  B {\bf 356}, 45 (1995);
%  [arXiv:hep-ph/9504292].
  %%CITATION = PHLTA,B356,45;%%
P.~Binetruy, S.~Lavignac and P.~Ramond,
  %``Yukawa textures with an anomalous horizontal Abelian symmetry,''
  Nucl.\ Phys.\  B {\bf 477}, 353 (1996);
%  [arXiv:hep-ph/9601243].
  %%CITATION = NUPHA,B477,353;%%
N.~Irges, S.~Lavignac and P.~Ramond,
  %``Predictions from an anomalous U(1) model of Yukawa hierarchies,''
  Phys.\ Rev.\  D {\bf 58}, 035003 (1998);
%  [arXiv:hep-ph/9802334].
  %%CITATION = PHRVA,D58,035003;%%
N.~Maekawa,
  %``Neutrino masses, anomalous U(1) gauge symmetry and doublet-triplet
  %splitting,''
  Prog.\ Theor.\ Phys.\  {\bf 106}, 401 (2001).
%  [arXiv:hep-ph/0104200].
  %%CITATION = PTPKA,106,401;%%




%\cite{Dreiner:2003yr}
\bibitem{Dreiner:2003yr}
  H.~K.~Dreiner, H.~Murayama and M.~Thormeier,
  %``Anomalous flavor U(1)X for everything,''
  Nucl.\ Phys.\  B {\bf 729}, 278 (2005).
%  [arXiv:hep-ph/0312012].
  %%CITATION = NUPHA,B729,278;%%

%\cite{Harnik:2004yp}
\bibitem{Harnik:2004yp}
  R.~Harnik, D.~T.~Larson, H.~Murayama and M.~Thormeier,
  %``Probing the Planck scale with proton decay,''
  Nucl.\ Phys.\  B {\bf 706}, 372 (2005);
%  [arXiv:hep-ph/0404260].
  %%CITATION = NUPHA,B706,372;%%
H.~K.~Dreiner, C.~Luhn, H.~Murayama and M.~Thormeier,
  %``Baryon Triality and Neutrino Masses from an Anomalous Flavor U(1),''
  Nucl.\ Phys.\  B {\bf 774}, 127 (2007);
%  [arXiv:hep-ph/0610026].
  %%CITATION = NUPHA,B774,127;%%
%``Proton Hexality from an Anomalous Flavor U(1) and Neutrino Masses - Linking
  %to the String Scale,''
  Nucl.\ Phys.\  B {\bf 795}, 172 (2008).
%  [arXiv:0708.0989 [hep-ph]].
  %%CITATION = NUPHA,B795,172;%%



%\cite{Gogoladze:2007ck}
\bibitem{Gogoladze:2007ck}
  I.~Gogoladze, C.~A.~Lee, T.~Li and Q.~Shafi,
  %``Fermion masses and mixings in GUTs with non-canonical U(1)Y,''
  Phys.\ Rev.\  D {\bf 78}, 015024 (2008).
%  [arXiv:0704.3568 [hep-ph]].
  %%CITATION = PHRVA,D78,015024;%%

%%%%%%%%%%%%%%%%%%%%%%%%%%%%%%%%%%%%%%%%%%%%%%%%%%%%%%%%%%%%%%%

%%%%%%%%%%%%%%%%%%%%%%%%%%%%%%%%%%%%%%%%%%%%%%%%%%%%%%%%%%%%%%%



%\cite{Ibanez:1991hv}
\bibitem{Ibanez:1991hv}
  L.~E.~Ibanez and G.~G.~Ross,
  %``Discrete gauge symmetry anomalies,''
  Phys.\ Lett.\  B {\bf 260} (1991) 291;
  %%CITATION = PHLTA,B260,291;%%
%``Discrete Gauge Symmetries And The Origin Of Baryon And Lepton Number
  %Conservation In Supersymmetric Versions Of The Standard Model,''
  Nucl.\ Phys.\  B {\bf 368}, 3 (1992).
  %%CITATION = NUPHA,B368,3;%%

%\cite{Dreiner:2005rd}
\bibitem{Dreiner:2005rd}
  H.~K.~Dreiner, C.~Luhn and M.~Thormeier,
  %``What is the discrete gauge symmetry of the MSSM?,''
  Phys.\ Rev.\  D {\bf 73}, 075007 (2006).
%  [arXiv:hep-ph/0512163].
  %%CITATION = PHRVA,D73,075007;%%

%\cite{Cvetic:1997ky}
\bibitem{Cvetic:1997ky}
  M.~Cvetic, D.~A.~Demir, J.~R.~Espinosa, L.~L.~Everett and P.~Langacker,
  %``Electroweak breaking and the mu problem in supergravity models with an
  %additional U(1),''
  Phys.\ Rev.\  D {\bf 56}, 2861 (1997)
  [Erratum-ibid.\  D {\bf 58}, 119905 (1998)].
%  [arXiv:hep-ph/9703317].
  %%CITATION = PHRVA,D56,2861;%%
P.~Langacker and J.~Wang,
  %``U(1) -prime symmetry breaking in supersymmetric E(6) models,''
  Phys.\ Rev.\  D {\bf 58}, 115010 (1998).
%  [arXiv:hep-ph/9804428].
  %%CITATION = PHRVA,D58,115010;%%


%\cite{Lee:2008pc}
\bibitem{Lee:2008pc}
  H.~S.~Lee,
  %``Lightest U-parity Particle (LUP) dark matter,''
  Phys.\ Lett.\  B {\bf 663}, 255 (2008).
%  [arXiv:0802.0506 [hep-ph]].
  %%CITATION = PHLTA,B663,255;%%



%%%%%%%%%%%%%%%%%%%%%%%%%%%%%%%%%%%%%%%%%%%%%%%%%%%%%%%%%%%%%%%

%%%%%%%%%%%%%%%%%%%%%%%%%%%%%%%%%%%%%%%%%%%%%%%%%%%%%%%%%%%%%%%

%\cite{Edsjo:1997bg}
\bibitem{Edsjo:1997bg}
  J.~Edsjo and P.~Gondolo,
  %``Neutralino Relic Density including Coannihilations,''
  Phys.\ Rev.\  D {\bf 56}, 1879 (1997);
%  [arXiv:hep-ph/9704361].
  %%CITATION = PHRVA,D56,1879;%%
%\cite{Niessen:2008hz}
%\bibitem{Niessen:2008hz}
  I.~Niessen,
  %``Supersymmetric Phenomenology in the mSUGRA Parameter Space,''
  arXiv:0809.1748 [hep-ph].
  %%CITATION = ARXIV:0809.1748;%%



%\cite{Belanger:2006is}
\bibitem{Belanger:2006is}
  G.~Belanger, F.~Boudjema, A.~Pukhov and A.~Semenov,
  %``micrOMEGAs2.0: A program to calculate the relic density of dark matter  in
  %a generic model,''
  Comput.\ Phys.\ Commun.\  {\bf 176}, 367 (2007).
%  [arXiv:hep-ph/0607059].
  %%CITATION = CPHCB,176,367;%%


%\cite{Djouadi:2002ze}
\bibitem{Djouadi:2002ze}
  A.~Djouadi, J.~L.~Kneur and G.~Moultaka,
  %``SuSpect: A Fortran code for the supersymmetric and Higgs particle spectrum
  %in the MSSM,''
  Comput.\ Phys.\ Commun.\  {\bf 176}, 426 (2007).
%  [arXiv:hep-ph/0211331].
  %%CITATION = CPHCB,176,426;%%



%\cite{Navarro:1996gj}
\bibitem{Navarro:1996gj}
  J.~F.~Navarro, C.~S.~Frenk and S.~D.~M.~White,
  %``A Universal Density Profile from Hierarchical Clustering,''
  Astrophys.\ J.\  {\bf 490}, 493 (1997).
%  [arXiv:astro-ph/9611107].
  %%CITATION = ASJOA,490,493;%%

%\cite{Sjostrand:2006za}
\bibitem{Sjostrand:2006za}
  T.~Sjostrand, S.~Mrenna and P.~Skands,
  %``PYTHIA 6.4 Physics and Manual,''
  JHEP {\bf 0605}, 026 (2006).
%  [arXiv:hep-ph/0603175].
  %%CITATION = JHEPA,0605,026;%%


%\cite{Delahaye:2007fr}
\bibitem{Delahaye:2007fr}
  T.~Delahaye, R.~Lineros, F.~Donato, N.~Fornengo and P.~Salati,
  %``Positrons from dark matter annihilation in the galactic halo: theoretical
  %uncertainties,''
  Phys.\ Rev.\  D {\bf 77}, 063527 (2008).
%  [arXiv:0712.2312 [astro-ph]].
  %%CITATION = PHRVA,D77,063527;%%



%\cite{Baltz:1998xv}
\bibitem{Baltz:1998xv}
  E.~A.~Baltz and J.~Edsjo,
  %``Positron Propagation and Fluxes from Neutralino Annihilation in the Halo,''
  Phys.\ Rev.\  D {\bf 59}, 023511 (1998).
%  [arXiv:astro-ph/9808243].
  %%CITATION = PHRVA,D59,023511;%%



%\cite{Maurin:2001sj}
\bibitem{Maurin:2001sj}
  D.~Maurin, F.~Donato, R.~Taillet and P.~Salati,
  %``Cosmic Rays below Z=30 in a diffusion model: new constraints on propagation
  %parameters,''
  Astrophys.\ J.\  {\bf 555}, 585 (2001);
%  [arXiv:astro-ph/0101231].
  %%CITATION = ASJOA,555,585;%%
%\cite{Hisano:2005ec}
%\bibitem{Hisano:2005ec}
  J.~Hisano, S.~Matsumoto, O.~Saito and M.~Senami,
  %``Heavy Wino-like neutralino dark matter annihilation into antiparticles,''
  Phys.\ Rev.\  D {\bf 73}, 055004 (2006).
%  [arXiv:hep-ph/0511118].
  %%CITATION = PHRVA,D73,055004;%%


%\cite{Meade:2009iu}
\bibitem{Meade:2009iu}
  P.~Meade, M.~Papucci, A.~Strumia and T.~Volansky,
  %``Dark Matter Interpretations of the Electron/Positron Excesses after
  %FERMI,''
  arXiv:0905.0480 [hep-ph].
  %%CITATION = ARXIV:0905.0480;%%


\bibitem{bp}
 V. Barger and R. J. N. Phillips, ``Collider Physics'', 1996.



%%%%%%%%%%%%%%%%%%%%%%%%%%%%%%%%%%%%%%%%%%%%%%%%%%%%%%%%%%%%%%%%%%%%%%

%%%%%%%%%%%%%%%%%%%%%%%%%%%%%%%%%%%%%%%%%%%%%%%%%%%%%%%%%%%%%%%%%%%%%%




%%%%%%%%%%%%%%%%%%%%%%%%%%%%%%%%%%%%%%%%%%%%%%%%%%%%%%%%%%%%%%%%%%%%%%

%%%%%%%%%%%%%%%%%%%%%%%%%%%%%%%%%%%%%%%%%%%%%%%%%%%%%%%%%%%%%%%%%%%%%%


%%%%%%%%%%%%%%%%%%%%%%%%%%%%%%%%%%%%%%%%%%%%%%%%%%%%%%%%%%%%%%%%%%%%%%

%%%%%%%%%%%%%%%%%%%%%%%%%%%%%%%%%%%%%%%%%%%%%%%%%%%%%%%%%%%%%%%%%%%%%%


\end{thebibliography}
\end{document}